\documentclass[12pt]{article}

\renewcommand{\a}{\alpha}
\renewcommand{\b}{\beta}

\renewcommand{\d}{\delta}

\newcommand{\h}{\eta}
\renewcommand{\th}{\theta}

\renewcommand{\l}{\lambda}
\renewcommand{\L}{\Lambda}
\newcommand{\m}{\mu}
\newcommand{\n}{\nu}
\newcommand{\p}{\pi}
\renewcommand{\P}{\Pi}
\renewcommand{\r}{\rho}

\renewcommand{\S}{\Sigma}
\renewcommand{\t}{\tau}

\newcommand{\f}{\phi}

\newcommand{\x}{\chi}

\newcommand\ba{\begin{array}}
\newcommand\ea{\end{array}}
\newcommand{\be}{\begin{enumerate}}
\newcommand{\ee}{\end{enumerate}}
\newcommand{\bi}{\begin{itemize}}
\newcommand{\ei}{\end{itemize}}
\newcommand{\bd}{\begin{description}}
\newcommand{\ed}{\end{description}}
\newcommand{\beq}{\begin{equation}}
\newcommand{\eeq}{\end{equation}}
\newcommand{\beqa}{\begin{eqnarray}}
\newcommand{\eeqa}{\end{eqnarray}}

\newcommand{\ld}{\ldots}
\newcommand{\cd}{\cdots}
\newcommand{\mr}{\mathrm}
\newcommand{\op}{\oplus}
\newcommand{\bop}{\bigoplus}
\newcommand{\ot}{\otimes}

\newcommand{\mc}{\mathcal}

\newcommand{\8}{\infty}
\newcommand{\ra}{\rightarrow}

\newcommand{\tb}{\textbf}
\newcommand{\pa}{\partial}
\newcommand{\mb}{\mathbf}

\newcommand{\sect}{\section}
\newcommand{\ssect}{\subsection}
\newcommand{\sssect}{\subsubsection}

\author{Smaragda Kessari}

\begin{document}

\begin{center}
                {\large\bf Affine Histories in Quantum Gravity:
                \vspace{0.3cm}
            Introduction and the Representation for a
            Cosmological Model
                           }
\end{center}

\vspace{0.7cm}

\begin{center}
Smaragda Kessari\footnote{s.kessari@ntlworld.com}
\end{center}

\begin{center}
May 2006
\end{center}

\vspace{0.7cm}

\begin{center}
The dreams of this work are dedicated to the memory of Nena,
Fouskas and Chestnut.
\end{center}

\vspace{1cm}

\begin{abstract}
It is shown how consistent histories quantum cosmology can be
realised through Isham's Histories Projection Operator consistent
histories scheme. This is done by using an affine algebra instead
of a canonical one and also by using cocycle representations. A
regularisation scheme allows us to find  a history Hamiltonian
which exists  as a proper self-adjoint operator. The role of a
cocycle choice is also discussed.
\end{abstract}

\sect{Introduction to the article}

The problem of time in quantum gravity  is  a particularly
complicated and difficult one to understand and deal with. This is
mostly because of the fact that general relativity being invariant
under the group of diffeomorphisms of the spacetime manifold
suggests a type of quantum physics that lacks any notion of time,
or at least any notion of time as we know it so far (for reviews
see \cite{CJIProbTime} and \cite{CJIQuestTime}). Various schemes
have been suggested to overcome this problem, one of which
involves uncovering an `internal' time from the existing variables
in the theory. In the present article we shall work with this
approach in a history context. More precisely, we will discuss the
history version of a Friedmann-Robertson-Walker (FRW) universe
coupled to a scalar field $\phi$, and in which we select (before
quantisation) the internal time to be this field $\f$.

Our main concerns are to apply consistent histories theory  to
quantum gravity and to find the correct representation of the
history algebra for quantum cosmology. We will argue that the
correct representation for quantum cosmology (with a history
Hamiltonian that is self-adjoint) involves using the history
analogue of the `affine group' that has been advocated as the
correct canonical group in normal canonical quantum gravity (for
an extensive presentation  see \cite{LesHouches}).

The main reason why the consistent histories approach to quantum
theory has been chosen by the author to tackle problems in quantum
gravity is the fact that in the consistent histories scheme the
conventional view of time need not play any fundamental role and
as a result---especially in the History Projection Operator
(`HPO') formalism (see later)---the scheme allows for logical
temporal connectives to be defined without prejudice about the
nature of time. In this way certain `conceptions' of time can
potentially be used in situations where there is no conventional
time.

This means that consistent histories might be used naturally and
profitably in a theory of quantum gravity where the notion of time
as an external parameter disappears. This is particularly true in
the case of quantum cosmology where one tries to apply quantum
theory to the entire universe. In addition,  the conventional
quantum-theory split of observer-system---with the associated
notions of an `external measurement' and `state vector
reduction'---is no longer appropriate in quantum cosmology. One of
the main features of the consistent history scheme is that ideas
of this type play no fundamental role. So that is another good
reason for using history ideas in quantum cosmology. We shall
mention here that an attempt to apply HPO histories to quantum
cosmology has already been made in \cite{AnastCosm}, but in a very
different way from that of this paper; in particular, these
authors did not use the `affine' history algebra that we shall
adopt.

In summary, we will show how the above choices lead to a
well-defined Hamiltonian operator in quantum cosmology. We find
the `history version' of a quantum cosmological model Hamiltonian
in which one uses a history analogue of the affine commutation
relations. Then we find a representation of the affine algebra in
which the, appropriately regularised, history Hamiltonian exists
as a proper renormalised self-adjoint operator.

Of course, this treatment challenges us to apply similar ideas to
the notion of space, something that has unfortunately not been
researched yet. However, the consistent histories formalism has
already generalised the Hilbert space used in quantum theory, even
when there was no original intention for that to happen. And one
would not be wrong to say that the consistent histories approach
allows for the removal of many established prejudices about what
is time, space, measurement, observer etc.\  and lets the theory
work without them.

The structure of the article is as follows. In chapter \ref{Intro}
we introduce the reader to the constituent theories and concepts
that will later be used to formulate the affine histories theory
and the application of the HPO scheme to quantum cosmology. First
the basics of consistent histories theory is introduced as it
first appeared in the version of Gell-Mann and Hartle \cite{G+H}.
Then the HPO theory is presented, with emphasis on the way in
which it can be developed to handle a continuous time parameter
\cite{CJI1}, \cite{CJI2}, \cite{CJI3}. In section \ref{affine} we
introduce the affine group of canonical quantum gravity
\cite{IKI}, \cite{IKII}. In section \ref{FRW} we show how the
original canonical quantisation of the FRW cosmological model
unintentionally leads to an `affine representation' although that
was not appreciated by the authors at that time.

In chapter \ref{AHT} the novel `affine histories' approach is
unravelled. First we just give the basic relations in quantum
gravity and then we show how that algebra is realised in quantum
cosmology. By doing this we can relate back to the earlier results
in \cite{BI} for the FRW model plus scalar field. Then we
introduce the history analogue of the affine relations and study
their representations using coboundaries in the exponential
Hilbert space of a simple, but very singular, representation.
Finally we introduce a Klauder-type regularisation of the singular
products in the definition of the history Hamiltonian. That choice
along with a choice of a cocycle lead to our conclusions about the
validity of the affine history~Hamiltonian.

\sect{Introduction to the main concepts} \label{Intro}

\ssect{Consistent histories theory}\label{CH}

\subsubsection{The emergence of histories in standard quantum
theory}\label{CHQT}

One way to introduce the histories theory is through the
conditional probability scheme of standard quantum theory which is
associated with the ideas of state-vector reduction. Suppose an
open, Hamiltonian quantum system is subjected to measurements by
an external (classical) observer. Let $\hat U(t_1,t_0)$ denote the
unitary time-evolution operator from time $t_0$ to time $t_1$. If
we let $\hat \r(t_0)$ denote the density operator of the quantum
system at time $t_0$, then at time $t_1$, in the Schr\"{o}dinger
representation, it will have evolved through a unitary
transformation to
            \beq
\hat{\r}(t_1)=\hat U(t_1,t_0)\hat \r(t_0)\hat
U(t_1,t_0)^{\dag}=\hat U(t_1,t_0)\hat \r(t_0)\hat U(t_1,t_0)^{-1}.
\eeq
 If at time $t_1$ we make a measurement of a property $\a_1$,
represented by a projection operator $\hat \a_1$, then the
probability that the property will be found is \beq \mr{Prob}(
\a_1 =1; \hat \r(t_1))=\mr{tr}(\hat \a_1 \hat
\r(t_1))=\mr{tr}(\hat \a_1(t_1) \hat \r(t_0)), \label{eq:Prob1}
\eeq where $\hat \a_1(t_1)$ is the Heisenberg-picture operator
defined as
            \beq
\hat \a_1(t_1):=\hat U(t_1,t_0)^{\dag}\hat \a_1(t_0)\hat
U(t_1,t_0). \eeq Then according to the Von-Neumann-L\"{u}ders
`reduction' postulate, the density matrix after the measurement is
(look, for example, in \cite{IshamQT} for more details)
            \beq
\hat \r_{\mr{red}}(t_1):=\frac{\hat \a_1(t_1)\r(t_0)\hat
\a_1(t_1)}{\mr{tr}(\hat \a_1(t_1)\hat \r(t_0))}. \eeq

If at time $t_2>t_1$ we perform another measurement on the same
system of a second property $\a_2$ (represented by the projection
operator $\hat\a_2$), then the conditional probability of getting
$\a_2=1$ at $t_2$ given that $\a_1=1$ at $t_1$ is
            \beq
\mr{Prob}(\a_2=1, t_2 | \a_1=1, t_1; \hat \r(t_0))=\mr{tr} (\hat
\a_2(t_2) \hat \r_{\mr{red}}(t_1))=\frac{\mr{tr}(\hat \a_2(t_2)
\hat \a_1(t_1)\hat \r(t_0)\hat \a_1(t_1))}{\mr{tr}(\hat
\a_1(t_1)\hat \r(t_0))}. \label{eq:Prob21} \eeq Then the
probability of getting $\a_1=1$ at $t_1$ \emph{and} $\a_2$ at
$t_2$ is (\ref{eq:Prob1})$\times$(\ref{eq:Prob21}), that is
            \beq
\mr{Prob}(\a_2=1, t_2 \ \mr{and}\  \a_1=1, t_1; \hat
\r(t_0))=\mr{tr} (\hat \a_2(t_2) \hat \a_1(t_1)\hat \r(t_0)\hat
\a_1(t_1)). \eeq Similarly, by performing an $n$-sequence of
measurements, the probability of finding \\ $\a_{1},\a_{2},
\ldots, \a_{n}$ at times $t_1,t_2, \ld, t_n$ is
            \beqa
&& \mathrm{Prob}(\a_{1}=1, t_1 \ \mathrm{and} \ \a_{2}=1, t_2 \
\mathrm{and}\ \ld \ \a_{n}, t_n  ;  \hat \r(t_0))
\nonumber \\
&& \ \ \ \ \  = \mathrm{tr}(\hat \a_{n}(t_n) \cd \hat
\a_{1}(t_1)\hat \r(t_0)\hat \a_{1}(t_1)\cd \hat
\a_{n}(t_n)).\label{Proban} \eeqa

In the consistent histories approach to quantum theory we call any
time-ordered sequence $\a:=(\hat \a _1, \hat \a_2,\ld, \hat \a_n)$
of projection operators  a \emph{homogeneous history}; the
associated set of time points $(t_1,t_2,\ld, t_n)$, with $t_1<
t_2<\cdots < t_n$, is known as the \emph{temporal support} of the
history. The main mathematical object of interest in this scheme
is the \emph{decoherence functional} $d_{\r}(\a,\b)$, which is
defined on pairs of homogeneous histories $\a,\b$ by
            \beq
d_{\r}(\a , \b):=\mr{tr}(\hat \a_n(t_n) \cd \hat \a_2(t_2)\hat
\a_1(t_1)\hat \r(t_0)\hat \b_1(t_1)\hat \b_2(t_2) \cd \hat
\b_n(t_n))=\mr{tr}(C_{\hat \a} \r(t_0)C_{\hat \b}^{\dag}),\eeq
with
            \beq
C_{\a}:=\hat \a_n(t_n) \cd \hat \a_2(t_2)\hat
\a_1(t_1)=U(t_0,t_n)\hat \a_n U(t_n,t_{n-1})\cd \hat
\a_2U(t_2,t_1)\hat \a _1 U(t_1,t_0)\label{dec} \eeq where $\hat
\a_i(t_i):=U(t_i,t_0)^{\dag}\hat \a_i U(t_i,t_0)$ is the
Heisenberg picture operator defined with respect to the fiducial
time $t_0$. The decoherence functional is defined for a closed
system, with no external observers and state-vector reductions; it
measures the interference between two histories that have some
properties at particular times and no reference to measurements is
needed. The crucial assumption of the history formalism is that,
under certain circumstances, the probability (\ref{Proban})
\emph{is} meaningful even in the absence of explicit measurements.
The way this is realised is by requiring that $d_{\r}(\a, \b)=0$
for all pairs $\a,\b$, $\a\neq\b$, in a set of histories. Such a
set of histories is said to be \emph{consistent}.

In this article we will only be interested in the appropriate
representation of the histories $ \a$ and (an analogue of) the
`class operators' $C_{\a}$, leaving the very important subject of
the decoherence functional for a later publication. However, it is
important to mention some features of the decoherence functional
to aid the comprehension of the histories theory and, in
particular, its incorporation in quantum cosmology.

One of these is that the decoherence functional encodes all the
information about the history---both its dynamical structure and
its initial state---whereas the history itself is just an ordered
(time-ordered) sequence of projection operators, each one of which
is a `proposition' (in the language of quantum logic) that in
standard quantum theory refers to the results of measurements at
times $t_1, t_2, \ld , t_n$ where $t_1< t_2<\ld <t_n$. Let us also
clarify here that a proposition asserts the value of some
observable in some range of the real numbers at a given time. In
our example a history $ \a =( \hat\a _1, \hat\a_2, \ld ,
\hat\a_n)$ is the sequential conjunction ``$\a_1$ is true at time
$t_1$, and then $\a_2$ is true at time $t_2$, and then $\ld$, and
then $\a_n$ is true at time $t_n$''.

\sssect{The Gell-Mann and Hartle approach}\label{GH}

In order to understand the approach to history theory that is
followed in this article it is useful to describe how it arose
originally. The consistent histories approach to quantum theory
was discovered independently by Omn\`{e}s, Griffiths, and
Gell-Mann and Hartle (GH) \cite{OmGr}, \cite{G+H}. It has become
an area of particular interest in the foundations of quantum
theory, and it has provided a new way of both realising and
interpreting quantum gravity with its emphasis on the fact that
everything that can be said about the physical world (including
its classical aspects) can be expressed in terms of sets of
decohering histories.

Two main schools in the field of decohering histories are the
path-integral approach, followed mainly recently by Hartle and
Halliwell \cite{H+H}, and the HPO approach: a generalisation of
the GH approach that was discovered and developed by Isham, Linden
and Schreckenberg \cite{CJI1}, \cite{CJI2}. Here we will only
mention the developments that will be relevant to the present
article.

Gell-Mann and Hartle made two important suggestions. The first one
was that the histories formalism can be extended to include
disjoint\footnote{Two homogeneous histories $\a:=(\hat \a_1,\hat
\a_2, \ld, \hat \a_n)$ and $\b:=(\hat \b_1,\hat \b_2, \ld, \hat
\b_n)$ are \emph{disjoint} if for at least one time point $t_i$ ,
$\hat\b_i$ is disjoint from $\hat \a_i$, i.e. the ranges of these
two projection operators are orthogonal subspaces of $\mc{H}$.}
\emph{sums} of homogeneous histories, known as
\emph{inhomogeneous} histories; this was done by introducing a
mechanism for forming a logical `or' for disjoint histories as
well as a logical `not'.

The second suggestion of GH was that a history could be regarded
as a fundamental quantity in its own right and not be just thought
of as a time-ordered sequence of projection operators/single-time
propositions. They achieved this by introducing a set of axioms
that were later reformulated and extended by Isham in order to
broaden and give a more solid mathematical foundation to the
theory. We shall discuss this shortly.

\sssect{Histories Projection Operator (HPO)
 initial conception}\label{HPO}

The work of Gell-Mann and Hartle gave the history theory a new
perspective with new potentials for future research. However, to
realise these fully, a new mathematical framework was necessary,
and this was provided by Isham \cite{CJI1}, and Isham and Linden
\cite{CJI2} in the form of the `HPO theory', the consistent
histories theory that is used here.

The key step was to introduce ideas of sequential/temporal quantum
logic and to represent such propositions by replacing the class
operator of GH---which is not a projection operator---with a
\emph{tensor} product of projectors---which \emph{is} a projection
operator. Note that, until the invention of the HPO formalism,
almost all studies of quantum logic had involved propositions at a
single time, and were therefore not applicable to the histories
theory.

As far as the use of tensor products is concerned, let us first
recall that in the GH approach, a (homogeneous) history $\a$ is a
set $(\hat\a_1, \hat\a_2, \ld,  \hat\a_n)$ of single-time
propositions which is represented by the class operator
$C_{\a}:=\hat \a_n \cd \hat \a_2\hat \a_1$. The product of
projectors is in general not itself a projector (this would
require them all to commute pairwise), and hence by using $C_\a$
the link with quantum logic is lost. However, the key observation
of Isham \cite{CJI1} was that a homogeneous history can also be
represented by the tensor product $\hat D_\a:=\hat \a_n \ot \hat
\a_{n-1} \ot \cd \ot \hat \a_1$  which is defined on the
\emph{tensor} product of $n$ copies of the single-time Hilbert
space $\mc{H}$. Unlike $C_\a$, the new quantity $\hat D_\a$
\emph{is} a projector.

This development fitted in very well with the suggestion by
Gell-Mann and Hartle that history theory could be suitable for
quantum gravity where the classical notions of space and time are
not applicable. As a result, as long as propositions about the
analogue of `histories' exist in a theory of this type, it becomes
reasonable to look for their representation with projectors on a
new `history' Hilbert space.

The HPO scheme has some attractive features. For example, by
applying the usual logical operations on projection operators, the
space of history propositions can be identified with an
orthoalgebra, or lattice\footnote{The notable difference between
those two is that in the former  the `or' proposition is defined
only on pairs of disjoint elements whereas in a lattice it is
defined on any pair (therefore, an orthoalgebra seems to be the
minimal structure that can be used in quantum logic).}, of history
propositions. That way, the set of propositions about the
histories becomes equipped with the structure of a
non-distributive lattice, and the lattice of projection operators
in a suitable Hilbert space can model this generalised-history
space.

More precisely, the proposition `$\a_1$ at time $t_1$' can be
represented on the two-time history Hilbert space by the operator
$\hat 1 \ot \hat \a_1$, and similarly the proposition `$\a_2$ at
time $t_2$' can be represented by the operator $\hat \a_2 \ot \hat
1$. Then the sequential conjunction `$\a_1$ at time $t_1$
\emph{and then} $\a_2$ at time $t_2$' is represented by $\hat
1\ot\hat \a_1 \wedge \hat\a_2\ot\hat 1$, which is equal to
$\hat\a_2 \ot \hat \a_1$, the HPO representation of the history
$(\hat \a_1, \hat \a_2)$. Therefore, a homogeneous history can be
identified with a sequential conjunction, which is in agreement
with its intrinsic and logical meaning.

\sssect{Continuous time and the history group in generalised
quantum theory and quantum field theory} \label{HPOcont}

As our interest is centred in quantum cosmology, the question now
arises about how the history formalism can be extended from a
finite temporal support to a continuous one and so we need to
consider what meaning can  be given to `continuous' temporal
logic. There have been two approaches to this question by Isham
and Linden \cite{CJI2}, and by Isham, Linden et al \cite{CJI3}.
Both of them have the same starting point, which is the
\emph{history group}: a history analogue of the canonical group
and the associated canonical commutation relations (CCR) used in
single-time quantum theory. This will play a central role in our
discussion of quantum cosmology.

The Lie algebra of the Weyl-Heisenberg group of transformations of
classical state-space gives the classical observables that can be
represented by self-adjoint operators in the associated quantum
theory, with the classical Poisson brackets going to operator
commutators. For example, consider the quantum theory of a
particle moving on the real line $\mathbf{R}$. The Lie algebra of
the Weyl-Heisenberg group is\footnote{We set $\hbar$=1 from now
on.}
            \beq \label{CCR}
[\hat x, \hat p]=i \eeq and what one looks for is an irreducible,
unitary representation of this group. Then, according to the Stone
and von Neumann theorem, the familiar representation on wave
functions $\psi(x)$ is unique up to unitary isomorphisms.

What is sought for in the HPO analogue is a unitary representation
of a \emph{history group} with a \emph{history algebra} whose
self-adjoint representations will give the projection operators
that will represent the propositions about the continuous-time
histories of the system.

As discussed in the previous section, a homogeneous history
$\a=(\hat \a_1, \hat\a_2, \ld, \hat \a_n)$ is represented by the
projection operator $\hat\a_n \ot \hat\a_{n-1}\ot \cd \ot \hat
\a_1$ that acts on the tensor product
$\mc{V}_n:=\mc{H}_n\ot\mc{H}_{n-1} \ot \cd \ot \mc{H}_1$ of
$n$-copies of the Hilbert space $\mc{H}$ of the canonical theory.
Since each $\mc{H}$ has a representation of the Lie algebra
(\ref{CCR}), the space $\mc{V}_n=\mc{H}_n\ot\mc{H}_{n-1} \ot \cd
\ot \mc{H}_1$ carries a representation of the Lie algebra
        \beq [\hat x_n \ot \hat 1
\ot\cd\ot \hat1, \hat x_n \ot \hat 1 \ot\cd\ot \hat1]=[\hat x_n
\ot \hat 1 \cd\ot \hat1,\hat1 \ot \hat x_{n-1} \ot \hat 1 \cd \ot
\hat1]=\ld =0 \eeq
        \beq
[\hat p_n \ot \hat 1\ot  \cd\ot \hat1,\hat p_n \ot \hat 1
\ot\cd\ot \hat1]=[\hat p_n \ot\hat1\ot \cd\ot \hat1,\hat1 \ot \hat
p_{n-1} \ot\cd\ot \hat 1  ]=\ld =0 \eeq
        \beq
[\hat x_n \ot\hat1 \cd\ot \hat1,\hat p_n \ot\hat1 \cd \ot\hat1]=[
\hat 1 \ot\hat x_{n-1} \ot\cd\ot \hat1,\hat1 \ot \hat p_{n-1} \ot
  \cd \ot\hat1]=\ld=i \eeq
        \beq
[\hat x_n \ot\hat1 \cd\ot \hat1,\hat 1 \ot \hat p_{n-1} \ot \cd
\ot\hat1]=[ \hat 1 \ot\hat x_{n-1} \ot\cd\ot \hat1,\hat p_n
\ot\hat 1\ot \cd \ot\hat1]=\ld=0 \eeq etc. More succinctly, we
have
            \beqa
&& \label{HCR1} [\hat x_{k},\hat p_{m}]=i\delta_{km}\\
&& \label{HCR2} [\hat p_{k},\hat p_{m}]=0\\
&& \label{HCR3} [\hat x_{k},\hat x_{m}]=0 \eeqa
 with
$k,m=1,2,\ldots,n$, where all the operators $\hat x_{k}$ and $\hat
p_{m}$ are viewed as acting on the one Hilbert space
$\mathcal{V}_{n}$ of the tensor product of $n$-copies of the
Hilbert space $\mathcal{H}\simeq L^2(\mathbf{R})$ of the canonical
theory.

However, what is used in practice, especially in continuous-time
histories, is the reverse argument; {\em i.e.}, one starts from
(\ref{HCR1})--(\ref{HCR3}) as defining the history algebra, and
then seeks a representation of it. The Stone-Von Neumann theorem
shows that there is an essentially unique representation space on
$\mathcal{V}_{n}$. Therefore, each history corresponds to a
proposition about the values of position and momentum (or linear
combinations of them) at the appropriate times in the temporal
support, and is constructed using the spectral projectors of the
representation of the history Lie algebra.

In order to construct the continuous-time histories, we suppose
that the analogue of (\ref{HCR1})--(\ref{HCR3}) is the algebra
\beqa && \label{HCCR1} [\hat x_{t_1},\hat
p_{t_2}]=ic_{[t]}\delta(t_{1}-t_{2})\\
&& \label{HCCR2} [\hat p_{t_1},\hat p_{t_2}]=0\\
&& \label{HCCR3} [\hat x_{t_1},\hat x_{t_2}]=0 \eeqa where
$-\infty \leq t_{1},t_{2} \leq \infty$ and $c_{[t]}$ is a
dimensional constant which has been introduced to make dimensions
on both sides of the equations agree \cite{CJI3}. We note that
this algebra is infinite dimensional, as can be seen from the fact
that these equations resemble those found in one-dimensional
quantum field theory.

Introducing a test-function space, which we assume is some (dense)
subset of the space $L^{2}(\mathbf{R})$ of real, square-integrable
functions on $\mathbf{R}$, we can introduce the smeared quantities
$\hat x_{f}=\int dt\, f(t) \hat x_t$ and $\hat p_{g}= \int dt\,
g(t) \hat p_t$ in terms of which the continuous history algebra
(CHA) becomes \cite{CJI3}
    \beqa \label{HCCRsm} {[}\hat x_{f},\hat p_{g}{]}&=&ic_{[t]}\int
f(t)g(t)\, dt = ic_{[t]}(f,g)\\
    {[}\hat x_{f_1},\hat x_{f_2}{]}&=&0\\
    {[}\hat p_{g_1},\hat p_{g_2}{]}&=&0 .
    \eeqa
where $(f,g):=\int f(t)g(t)\, dt$.

There are two different approaches one can take in order to find
the appropriate representation of the history algebra
(\ref{HCCR1})--(\ref{HCCR3}). In the first approach, \cite{CJI2},
one tries directly to construct the continuous analogue of a
tensor product. This involves using using coherent states
\cite{CJI2} and leads to the familiar Bosonic Fock space in the
guise of an exponential Hilbert space (see below). One can then
show that the history group algebra (\ref{HCCR1})--(\ref{HCCR3}),
or the smeared form (\ref{HCCRsm}), has a natural representation
on this space.

The second approach, \cite{CJI3} starts with the history group,
with history algebra (\ref{HCCR1})--(\ref{HCCR3}), and studies the
representations of this infinite-dimensional algebra. The Stone
von Neumann theorem does not work in infinite dimensions, and in
fact there are infinitely many unitarily inequivalent
representations. Guided by the results in an old QFT paper of
Araki \cite{Araki}, one seeks a representation in which the
operator that represents history propositions about time-averaged
values of energy genuinely exists (Araki showed that, in normal
quantum field theory, the requirement that the Hamiltonian for a
free field exists, uniquely determines the representation).

We will not discuss any details here of the original papers on
continuous history theory as it turns out that the history
analogue of the affine algebra has a representation theory that is
radically different from that associated with the canonical
history algebra in (\ref{HCCR1})--(\ref{HCCR3}). However, as we
shall see, it is still useful to employ the language of Fock
spaces and exponential Hilbert spaces, although these entities now
arise in a very different way.

However, the fact that the basic history affine algebra
representations are very different from those of
(\ref{HCCR1})--(\ref{HCCR3}) is not the only reason why the
representation methods in \cite{CJI3} cannot be applied to the
affine case. Another major reason is that, in \cite{CJI3}, a key
role is played by the fact that the Hamiltonian operator along
with the configuration variables $\hat x_t$ and $\hat p_t$ form a
(closed) Lie algebra. For if one finds the unitary representations
of the associated extended Lie group, then one guarantees the
existence of a self-adjoint Hamiltonian. Indeed, this is the
scheme used to \emph{construct} the Hamiltonian.

However, as we shall see, in our quantum cosmology model if one
adds the Hamiltonian to the set of the affine variables then the
resulting collection no longer forms a Lie algebra (this is
already true at the Poisson algebra level). As a result, in the
affine case one has to try to construct the Hamiltonian directly
(from the `outset') as a function of the affine history variables.
This is a tricky process involving regularisation and is the
procedure which we will follow\footnote{Actually, this phenomenon
arises already in normal quantum theory since, apart from the case
of the simple harmonic oscillator, the Hamiltonian
$\frac{p^2}{2m}+V(x)$ does not combine with $x$ and $p$ to give a
Lie algebra. In the field theory analogue, it is only the
Hamiltonian of the free field that combines with the other
variables to give a Lie algebra.}.

Finally, in order to conceptualise and understand the application
of HPO histories theory in cosmology, let us say a few parenthetic
words about the application of the CHA in quantum field theory by
Isham, Linden et al \cite{CJI3}. We foliate a four-dimensional
Minkowski space-time with the aid of a timelike vector $n^{\m}$,
normalised by $\h_{\m\n}n^{\m}n^{\n}=1$ where
$\h_{\m\n}=(+,-,-,-)$. The equal time CCR for bosonic quantum
field theory are
            \beq
[\f({\underline x}_1),\f(\underline x_2)]=0=[\p(\underline
x_1),\p(\underline x_2)]\eeq
            \beq
[\f(\underline x_1),\p(\underline x_2)]=i \d^3(\underline
x_1-\underline x_2). \eeq Following the same steps as we did when
going from (\ref{CCR}) to (\ref{HCCR1})--(\ref{HCCR3}), we
`historise' the quantum field CCR above and we get
            \beq \label{HQFTCR1}
[\f_{t_1}({\underline x}_1),\f_{t_2}(\underline
x_2)]=0=[\p_{t_1}(\underline x_1),\p_{t_2}(\underline x_2)]\eeq
            \beq  \label{HQFTCR2}
[\f_{t_1}(\underline x_1),\p_{t_2}(\underline x_2)]=ic_{[t]}
\d(t_1-t_2) \d^3(\underline x_1-\underline x_2). \eeq One can also
rewrite (\ref{HQFTCR1})---(\ref{HQFTCR2}) in a `four-vector' mode
as
            \beq
[\f(X), \f(Y)]=0=[\p(X), \p(Y)]\eeq
            \beq
[\f(X), \p(Y)]=ic_{[t]} \delta^{4}(X-Y) \eeq which however are
\emph{not} covariant commutation relations, as the four-vector
$x_n$ with $(t,x) \in \mathbf{R}\times \mathbf{R}^3$ (with $n\cdot
x_n=0$) of the covariant theory is related to the space-time point
$X$ with $X=tn+x_n$.

\ssect{The affine algebra in quantum theory and quantum
gravity}\label{affine}

\sssect{The motivation for using affine commutation relations}
\label{ACR} In \cite{LesHouches}, and later in \cite {IKI} and
\cite{IKII}, it was extensively and explicitly shown that although
the familiar CCR are appropriate for a system whose classical
configuration space, $Q$, is a vector space, this canonical
algebra is not appropriate for systems where $Q$ has a different
topology. However, group theory ideas \emph{can} still be employed
effectively provided that the configuration space is a homogeneous
space: \emph{i.e.}, $Q$ has some transitive group, $G$, of
transformations, so that $Q\simeq G/H$ for some subgroup $H$ of
$G$. The canonical group then becomes the semi-direct product
$W\times_s G$ where $W$ is a vector space carrying a linear
representation of $G$ and such that $Q$ is embedded equivariantly
as a $G$ orbit \cite{LesHouches}. This new canonical group
$W\times_s G$ is often referred to as an `affine' group.

The use of such group-theoretic methods becomes particularly
important when the configuration space $Q$ is
infinite-dimensional. Experience with standard QFT suggests that
the state vectors will be functionals on a `distributional dual'
of $Q$. However, it is not trivial to define what is meant by such
a dual when $Q$ is not a vector space (for example, what is meant
by a `distributional metric' in quantum gravity?) and the
group-theoretical approach to quantisation provides a powerful way
of addressing this question.

One simple, but important, example where one has to deal with a
non-vector space in quantum theory is the following. Suppose we
want to quantise a particle moving on the positive real line. The
standard CCR for $Q\simeq\mathbf{R}$ are $[\hat x, \hat p]=i$
which when exponentiated becomes $e^{-ia\hat p}\hat x e^{i a \hat
p}=\hat x -a$. But since $a$ is any real number, these CCR are
incompatible with the condition $x>0$. On the other hand, if we
instead take the \emph{affine} algebra (the `ACR')
            \beq \label{affQM}
[\hat x, \hat \p]=i\hat x, \eeq (where, classically, $\p=x p$),
and if we exponentiate this algebra, then we find that for any
(faithful) irreducible representation the spectrum of the operator
$\hat x$ is either the positive real numbers or the negative real
numbers. Indeed, as was first shown by Klauder
\cite{Kla70a70b73a73b} and then by Pilati \cite{Pil8283}, there
exist two unitarily inequivalent, faithful, irreducible
representations of the ACR: one where the spectrum of the variable
$\hat x$ of the configuration space is strictly greater than zero,
and one where it is strictly less than zero. So, by choosing the
appropriate representation we are able to quantise the system in a
way that is compatible with the classical requirement that $x>0$.

Similarly, in canonical quantum gravity one wants the metric
$g_{ij}$ to be a positive determinant tensor with signature
$(+++)$. Therefore, the configuration space, and consequently the
unconstrained phase space of $(g_{ij}, \p^{ij})$, is not a vector
space, and the corresponding analogue of the affine algebra
(\ref{affQM}) is \beqa && \label{metricACR1a}
[\hat{g}_{ij}(x),\hat{\pi}^{k}_{\
l}(y)]=\frac{i}{2}(\hat{g}_{il}(x) \delta_{j}^{\ k}
+\hat{g}_{jl}(x)\delta_{i}^{\ k})\delta^{(3)}(x-y)\\
&& \label{metricACR2a} [\hat{\pi}_{i}^{\ j}(x),
\hat{\pi}_{k}^{\
l}(y)]=\frac{i}{2}(\hat{\pi}_{k}^{\ j}(x)\delta_{\
i}^{l}-\hat{\pi}_{i}^{\ l}(x)\delta_{\ k}^{j})\delta^{(3)}(x-y)\\
&& \label{metricACR3a} [\hat{g}_{ij}(x),\hat{g}_{kl}(y)]=0.
\eeqa

In the simple minisuperspace model of quantum cosmology that is
considered below, the variable $g_{ij}(x)$ becomes just the radius
$R$ of the universe, and one has to have an algebra that is
compatible with the structure of this reduced classical
configuration space. But this is nothing but the positive real
numbers and hence, as remarked above in the context of
(\ref{affQM}), the appropriate algebra is
         \beq \label{RWACR}
[\hat R, \hat \P]=i\hat R. \eeq

\sssect{The  representation of the ACR}\label{ACRrepn} As we will
be concerned with a history, minisuperspace analogue of
(\ref{metricACR1a})--(\ref{metricACR3a}) let us consider briefly
what is known about the representations of the latter algebra, as
discussed in \cite{IKII}.

The group associated with this algebra is the semi-direct product
$\mc{C}=C^{\8}(\Sigma, S(3,\mathbf{R})) \times_s
C^{\8}(\Sigma,GL^{+}(3,\mathbf{R}))$ where $S(3,\mathbf{R})$ is
the (vector) space of $3\times 3$ real symmetric matrices and
$\Sigma$ is the spatial manifold. When used in canonical quantum
gravity, the task is to find unitary representations of this
group. Such a unitary representation automatically gives rise to
the self-adjoint operators that represent the associated physical
variables. From a technical perspective, we can  take advantage of
the fact that unitary operators are always bounded, unlike the
(smeared) operators $\hat g(v)$ and $\hat \p(M)$.

One way of finding a representation of the semi-direct product
group $\mc{C}$ is through the theory of induced representations
and Mackey's theorems \cite{Mac}. However, in such a case
\emph{all} the representations found in this way are exhaustive
only in the case where $\mc{C}$ is a finite-dimensional Lie group.
Nevertheless, in \cite{IKII} some limited results were obtained by
treating the infinite-dimensional case as a generalisation of the
finite-dimensional one. We shall now briefly discuss some of these
results.

If $W$ is a function space $C^{\8}(\S,V)$, with $V$ a
finite-dimensional vector space (in our example we have $V\simeq
S(3,\mathbf{R})$), its (topological) dual space,
$C^{\8}(\S,V)^{'}$, will play an important role in the quantum
theory. Indeed, general spectral theory tells us that, in general,
the state vector can always be written as a (possibly vector-space
valued) functional $\Psi[\chi]$ of $\chi\in C^{\8}(\S,V)^{'}$.

This space, $C^{\8}(\S,V)^{'}$, is some set of distributions, and
if $\chi\in C^{\8}(\S,V)^{'}$ we follow convention and write,
rather heuristically, for all $v\in C^{\8}(\S,V)$,
            \beq
\x(v)=\int_{\S}\x_{ij}(x)v^{ij}(x)d\th(x) \eeq where $d\th$ is the
measure on the three-manifold $\S$ associated with some background
volume element\footnote{The use of a background volume element can
be avoided if one employs densities of an appropriate weight as
the test functions $v$.}.

In our case, the simplest example of an element of the
distributional dual is the Dirac $\d$-function object,
$\d_{(x,g)}$, which is defined for each $x\in \Sigma$ and $g\in
GL^+(3,\mathbf(R))$, as
            \beq \label{chiijx}
    \d_{(x,g)}(v):=g_{ij}v^{ij}(x)\eeq
for all $v\in C^{\8}(\Sigma, S(3,\mathbf{R}))$. Thus we can write
    \beq
    \d_{(x,g)ij}(\cdot):=g_{ij}\d_x(\cdot)\label{chiijx2}
    \eeq where $(\cdot)$ reminds us that this is a generalised
function on $\Sigma$.

One of the major problems to be dealt with is the existence and
properties of the measure $\m$ that needs to be defined on the
quantum state domain space $C^{\8 }(\S,S(3,\mathbf{R}))^{'}$. This
is used to give the inner product on the quantum Hilbert space,
and it must be chosen so that the representation of the group
$\mc{C}$ is unitary. However, the space $C^{\8
}(\S,S(3,\mathbf{R}))^{'}$ is infinite-dimensional, and finding
measures of this type is very difficult. On the other hand, the
space of distributions of the type in (\ref{chiijx}) can be
identified with the space $\S\times GL^{+}(3,\mathbf{R})$ via the
injection \beqa \S\times GL^{+}(3,\mathbf{R})&\ra&
C^{\8'}(\S,GL^{+}(3,\mathbf{R}))\nonumber \\
 (x,g)&\mapsto& g_{ij}\d_{x}(\cdot) \label{map_ed} \eeqa

On this (rather thin) subspace of distributions we can use the
measure $d\th \ot dH$ on $\S\times GL^{+}(3,\mathbf{R})$ where
$dH$ is the Haar measure on $GL^{+}(3,\mathbf{R})$. This enables a
representation of $\mc{C}=C^{\8}(\Sigma, S(3,\mathbf{R}))\times_s
C^{\8}(\Sigma,GL^{+}(3,\mathbf{R}))$ to be defined on
$L^{2}(\S\times GL^{+}(3,\mathbf{R}));d\th \ot dH)$ which, taking
into account the group law\footnote{The group law of this group
is, for all $x\in\Sigma$,
$(\L_1(x),v_1(x))(\L_2(x),v_2(x))=(v_1(x)+\L_1(x)v_2(x),\L_1(x)\L_2(x))$
where $\L_1,\L_2\in C^{\8}(\Sigma,GL^{+}(3,\mathbf{R}))$ and
$v_1,v_2 \in C^{\8}(\Sigma, S(3,\mathbf{R}))$.} and
$\d_{(x,g)}(v):=g_{ij}v^{ij}(x)$, is found to be
            \beqa
&& (V(v)\psi)(x,g)=\exp(i\S_{{i,j}=1}^3v^{ij}(x)g_{ij})\psi(x,g)
\label{VrepIKII}\\
&& (U(\L)\psi)(x,g)=\psi(x,\L^{-1}(x)g)\label{UrepIKII}. \eeqa We
note that this representation of $\mc{C}$ is reducible since state
functions $\psi(x,g)$ whose support in $x$ is some compact subset
$K\subset\mathbf{R}$ form an invariant subspace. This is a problem
to which shall return later in our discussion of quantum
cosmology.

By expanding (\ref{VrepIKII})--(\ref{UrepIKII}) one finds the
self-adjoint operators
            \beqa
&&(\hat g_{ij}(y)\psi)(x,g)=g_{ij}\d_{x}(y)\psi (x,g)
\label{erepIKII}\\
&& (\hat \p_{i}^{\ j}(y)\psi) (x,g)= -i\d_x(y)g_{ik}\partial \psi
/ \partial g_{jk}(x,g). \label{prepIKII}\eeqa This shows clearly
the sense in which the metric operator $\hat g_{ij}(x)$ (and the
associated momenta $\hat\p_i^{ \ j}(x)$) is concentrated on a
single space point. That gave the idea of generalising
(\ref{chiijx2}) to sums of distributions like
            \beq
\d_{(x_1,g^{(1)}; x_2,g^{(2)}; \ld ;
x_N,g^{(N)})}(\cdot):=\S_{n=1}^{N}g^{(n)}_{ij}\d_{x_{n}}(\cdot)
\eeq and (\ref{VrepIKII})--(\ref{UrepIKII}) are accordingly
replaced with
            \beqa
(V(v)\psi)(x_1,g^{(1)}; \ld ; x_N,g^{(N)})\!\!\!&:=& \!\!\! \exp (
i
\S_{n=1}^N\S_{i,j=1}^{3}v^{ij}(x_n)g^{(n)}_{ij})\psi(x_1,g^{(1)};
\ld ;
x_N,g^{(N)})  \label{zeus1}\ \ \ \ \ \ \  \\
(U(\L)\psi)(x_1,g^{(1)}; \ld ; x_N,g^{(N)})\!\!\!&:=&\!\!\!
\psi(x_1,\L^{-1}(x_1)g^{(1)}; \ld ; x_{N}, \L^{-1}
(x_{N})g^{(N)}).\label{zeus2}\ \ \ \ \ \ \ \eeqa

This representation too is reducible. However the possibility
arises of taking the direct sum (over $N$) of all such
representations and then introducing a cocycle \cite{EF} on the
associated exponential Hilbert space. This possibility was raised
in \cite{IKII}, and employed by Klauder \cite{KlaRep} and Pilati
\cite{Pil8283}. This suggestion was not developed in \cite{IKII}
but it is one that we shall explore in our history theory.

The most serious  problem with representations of the type in
(\ref{zeus1})--(\ref{zeus2}) is that they are very singular, which
causes big problems if one has to try to construct a product of
its operators, such as is needed (as we shall see in the following
subsection) for the Hamiltonian operator in our quantum
cosmological model. This is so first because the relevant product
is always calculated at the \emph{same} spatial point (spatial
point in the example above, but time point in the cosmological
model history case we shall see next) and secondly because of the
meaninglessness of a squared Dirac delta function. Klauder
\cite{Kla70a70b73a73b} suggested a regularisation of the product
of such representations which involves `dividing by' a delta
function $\delta(0)$ and we shall return to this later in the
context of quantum cosmology.

\ssect{A Friedmann-Robertson-Walker cosmological model}
\label{FRW}

In \cite{BI} the quantisation of the homogeneous and isotropic FRW
universe coupled to a scalar field was given. The metric is
            \beq \label{ds}
ds^2=N(t)^2 dt^2-R(t)^2 S_{ij}dx^i dx^j \eeq where $N(t)$ is the
lapse function (the normal component of the deformation
vector-hypersurface of constant time), $R(t)$ represents the
radius of the universe and $S_{ij}$ is the fixed background metric
for a three-space for the curvature constant $K$ taking any of the
values $1, 0, -1$ for a three-sphere, flat, or hyperbolic universe
respectively.

In the case of a massive scalar field the matter Lagrangian is
            \beq
\mc{L}=\frac{1}{2}(-\mr{det}g)^{1/2}(g^{\m\n}\pa_{\m}\f\pa_{\n}
\f-m^2\f^2) \eeq where $\f$ is the scalar field and $g_{\m\n}$ is
the four-metric of (\ref{ds}). As the spatial metric is
homogeneous, there are just three coupled variables: $N(t), R(t)$
and $\f(t)$.

 The ADM (Arnowitt, Deser, Misner) Lagrangian \cite{ADM} is
 \cite{BI}
\begin{equation}\label{eq:Lang1}
\mathcal{L}=\pi_{R}\dot{R}+\pi_{\phi}\dot{\phi}+
N\left(\frac{\pi_{R}^2}{24R}+6KR-R^3\phi^2 \ \frac{m^2}{2}-
\frac{\pi_{\phi}^2}{2R^3}\right)
\end{equation}
in which $N$ is a Lagrange multiplier, so that the quantity in the
parenthesis is set to $0$ by the   equations of motion. It is
common in canonical quantisation to choose one of the variables as
an `internal time'. In \cite{BI} various such  choices were made,
but here we shall adopt the one that is in line with our affine
scheme. This is the choice $\f=t$. Then, after solving the
constraint equation
\begin{equation}\label{eq:Ham1}
\frac{\pi_{R}^2}{24R}+6KR-R^3\phi^2 \ \frac{m^2}{2}-
 \frac{\pi_{\phi}^2}{2R^3}=0
\end{equation}
for $\pi_{\phi}$, and replacing it in the Lagrangian
(\ref{eq:Lang1}), we are led to the squared Hamiltonian
\begin{equation}
H^2=\frac{1}{12}R^2 \pi_{R}^2 +12KR^{4}-R^{6}t^2 m^2 .
\end{equation}

In trying to quantise this system, account must be taken of the
fact that, classically, $R$ is confined to lie in the interval
$(0, \infty)$. However, if we try to quantise the system in the
obvious way (with the wave-function vanishing at the end points)
using the usual assignments
\begin{equation}
R \rightarrow R \ , \ \ \ \ \pi_{R}\rightarrow -i \frac{d} {dR}
\end{equation}
satisfying the CCR
            \beq
[\hat R, \hat \p_R]=i \eeq then a problem arises because the
second operator turns out not to be self-adjoint (although it is
Hermitian)\cite{BI}.

In the original paper \cite{BI}, this problem was tackled in a
rather indirect way by first conjecturing a suitably quantized
Hamiltonian for the massless ($m=0$) and flat three-space ($K=0$)
case, \emph{i.e.}, for the case
\begin{equation}\label{eq:H2m0K0}
H^2=\frac{1}{12}R^2 \pi_{R}^2
\end{equation}
and showing that this has an interpretation as a positive
self-adjoint Hamiltonian whose positive square root exists. The
operator $\widehat{R\pi_R}$ was then defined as a combination of
$R$ and $\p_R$ that \emph{is} self-adjoint (and positive) and
specifically it was defined that the quantisation of $R^2 \p_R^2$
is  $\frac{d}{dR}R^2\frac{d}{dR}$ and then $\widehat{R\pi_R}$ was
defined to be the square root of this operator. However, we know
now from the previous discussion that $R \p_R$ is classically the
`affine conjugate' $\P$ of $R$, so that the classical Hamiltonian
can be written as
            \beq \label{Haffcosmo}
H^2=\frac{1}{12}\P^2+12KR^2-R^6t^2m^2  \eeq with the ACR being
             \beq
[\hat R, \hat \P]=i\hat R  \eeq as was discussed in Section
\ref{ACR}. This, physically meaningful, way explains the
mathematics and the structure behind the original way that was
successfully guessed by Blyth and Isham, and it offers the
possibility for further development and applications. The
Hamiltonian (\ref{Haffcosmo}) is the one whose history version we
are going to deal with below.

\sect{Affine histories theory}\label{AHT}

The logic for the steps that have already been followed was  (i)
start with the Lie algebra that one \emph{does} have {\em i.e.},
the one generated by the affine variables $R$ and $\Pi$; (ii)
construct the corresponding Lie group; and then (iii) look for the
unitary representations of this Lie group. In each such
representation the affine variables will arise as self-adjoint
operators representing the associated Lie algebra, and so the
final step is to construct a self-adjoint Hamiltonian in terms of
these affine variables.

We are interested in the history analogue of the ACR which, as we
will see, will involve variables $R(t)$ and $\Pi(t)$, and we may
start to worry that problems of divergences will arise when we
come to construct the history analogue of the Hamiltonian
(\ref{Haffcosmo}). The history analogue of Isham's or Klauder's
representations and regularisation methods for the affine algebra
might be of use here. If these representations do not work, then
new representations must be sought. However, we have already
argued that those representations are unlikely to be the correct
ones for an infinite-dimensional histories theory, and we shall
also see that that those regularisation methods can be replaced
with some more physical and efficient ones.

Let us briefly see (i) what \emph{has} been done already in terms
of the correct representation; (ii) what does not seem appropriate
to be followed; and (iii) what needs to be changed in an `affine
histories' scheme.

In \cite{CJI3} a representation of the CHA
(\ref{HCCR1})--(\ref{HCCR3}) was chosen for a particle moving on
the real line using the familiar ideas from quantum field theory,
especially the use of the bosonic Fock space as the history
Hilbert space. This employed the result that the (history analogue
of the) Hamiltonian operator and the history variables $x_t, p_s$
form a closed algebra. Then, since the time automorphism was
unitarily implementable, the corresponding self-adjoint operator
existed and could be identified as the time average of the energy
in the history theory. However, we have seen now that the CCR, and
consequently the associated history analogue, is not the correct
algebra for our quantum cosmology theory where $R>0$.

On the other hand in \cite{IKII}, and as discussed above, some
representations of the ACR in quantum gravity were found using
induced representation techniques. However, the metric operator
and its conjugate were defined on a single spatial point and an
effort to use an extension of that representation leads to its
further reducibility. However, it was suggested in \cite{IKII}
that the use of cocycles might change the situation and this
suggestion, in its history form, will be studied below.

\ssect{Affine history algebra in quantum cosmology}
\label{QGAffine}

Here, we shall not attempt to find a representation of the whole
quantum gravity affine history algebra. Rather, we shall study the
relevant history algebra for the quantum cosmological model of
interest. So, based on the arguments in \cite{CJI3} that were
analysed in Section \ref{HPOcont}, we claim that the history
version of the ACR (\ref{metricACR1a})--(\ref{metricACR3a}) is
            \beqa\label{eq:metricHACR1} && [\hat{g}_{ij}(x,t),
            \hat{\pi}^{k}_{\
l}(y,t')]=\frac{i}{2}c_{[t]} (\hat{g}_{il}(x,t) \delta_{j}^{\ k}
+\hat{g}_{jl}(x,t) \delta_{i}^{\ k})\delta^{(3)}(x-y)
\delta(t-t')\\
&& \label{eq:metricHACR2} [\hat{\pi}_{i}^{\ j}(x,t),
\hat{\pi}_{k}^{\ l}(y,t')]=\frac{i}{2}c_{[t]}(\hat{\pi}_{k}^{\
j}(x,t)\delta_{\ i}^{l}- \hat{\pi}_{i}^{\ l}(x,t)\delta_{\
k}^{j})\delta^{(3)}(x-y)\delta(t-t')\\
&& \label{eq:metricHACR3}
 [\hat{g}_{ij}(x,t),\hat{g}_{kl}(y,t')]=0
\eeqa which for a spatially homogeneous geometry becomes
\beqa\label{eq:metricHACR1homo}&&
[\hat{g}_{ij}(t),\hat{\pi}^{k}_{\
l}(t')]=\frac{i}{2}c_{[t]}(\hat{g}_{il}(t) \delta_{j}^{\ k}
+\hat{g}_{jl}(t)\delta_{i}^{\ k})\delta(t-t')\\
&& \label{eq:metricHACR2homo} [\hat{\pi}_{i}^{\ j}(t),
\hat{\pi}_{k}^{\ l}(t')]=\frac{i}{2}c_{[t]}(\hat{\pi}_{k}^{\
j}(t)\delta_{\ i}^{l}-\hat{\pi}_{i}^{\ l}(t)\delta_{\
k}^{j})\delta(t-t')\\
&& \label{eq:metricHACR3homo}
[\hat{g}_{ij}(t),\hat{g}_{kl}(t')]=0. \eeqa

To be mathematically well-defined, these operators must be smeared
with test functions, and we smear $\hat{g}_{ij}(t)$ with a density
field $f^{ij}(t)$, and $\hat{\pi}_{i}^{\ j}(t)$ (which is an
operator density) with an ordinary field $F^{i}_{\ j}(t)$. Then
the ACR (\ref{eq:metricHACR1homo}) become
\begin{equation}\label{eq: SRWHACR1}
[\hat{g}(f_{d}),\hat{\pi}(F_{o})]=\frac{ic_{[t]}}{2}\int dt \
\hat{g}_{il}(t)f^{ij}(t)F_{j}^{\ l}(t) + \frac{ic_{[t]}}{2}\int dt
\ \hat{g}_{jl}(t)f^{jl}(t)F_{i}^{\ l}(t)
\end{equation}
where $\hat{g}(f_{d})=\int dt \ \hat{g}_{ij}(t)f^{ij}(t)$, and
$\hat{\pi}(F_{o})=\int dt \ \hat{\pi}^{k}_{ \ l}(t)F_{k}^{\
l}(t)$; there is also an appropriate analogue for
(\ref{eq:metricHACR2homo}) and (\ref{eq:metricHACR3homo}). The
group of the ACR algebra (\ref{metricACR1a})--(\ref{metricACR3a})
is the semi-direct product
 ${C^{\8}(\Sigma, S(3,\mathbf{R}))\times_s
C^{\8}(\Sigma,GL^{+}(3,\mathbf{R}))}$, and similarly the group of
the algebra (\ref{eq:metricHACR1homo})--(\ref{eq:metricHACR3homo})
is $C^{\8}(T, S(3,\mathbf{R}))\times_s
C^{\8}(T,GL^{+}(3,\mathbf{R}))$ where $T$ denotes the `time-line'
which could be the whole real line $\mathbf{R}$ or it might be
some bounded subset of $\mathbf{R}$ according to the physical
situation under consideration.

Let us now consider the FRW cosmological model case analysed in
Section \ref{FRW}. It is easy to see that the history--ACR will
take the form
            \beqa\label{RWHACR1}
&&[\hat{R}(t),\hat{\P}(t')]=ic_{[t]}\hat{R}(t)\delta(t-t')\\
&& \label{RWHACR2} [\hat{\P}(t),\hat{\P}(t')]=0 \ ,  \ \
[\hat{R}(t),\hat{R}(t')]=0 \eeqa which in smeared form becomes
            \beqa
&& [\hat R(f_R),\hat \P(f_\P)]=ic_{[t]}\int dt\;
f_{R}(t)f_\P(t)\hat
R(t):= ic_{[t]} \hat R(f_R f_\P) \\
&& [\hat \P(f_\P),\hat \P(f'_\P)]=0\ ,  \ \  [\hat R(f_R), \hat
R(f'_R)]=0 \eeqa where $\hat R(f_R):=\int \hat R(t) f_R(t) dt$ and
$\hat \Pi(f_\Pi)=\int \hat \Pi(t) f_\Pi(t) dt$, etc. The
corresponding history group is $C(T,\mathbf{R})\times_{s}
C(T,\mathbf{R}_{+})\cong C(T,\mathbf{R}\times_{s}\mathbf{R}_{+})$,
{\em i.e.}, the group of all continuous maps from the time-line
$T$ to the group $\mathbf{R}\times_{s}\mathbf{R}_{+}$.

The history analogue of the Hamiltonian squared (\ref{Haffcosmo})
 is
            \beq \label{Hhistaffcosmo}
 \hat H^2(t)=\frac{1}{12}\hat \P^2(t)+12K \hat R^4(t)-\hat
 R^6(t)t^2m^2
 \eeq
and our task is to find a representation of the algebra in
(\ref{RWHACR1})--(\ref{RWHACR2}) in which $\hat H^2(t)$, or a
suitably smeared version of it, exists as a proper self-adjoint
operator.

Of course, the physically relevant quantity is $\hat H$ which will
be the square root of (\ref{Hhistaffcosmo}), and one immediately
obvious problem is that when $m\neq 0$, the operator in
(\ref{Hhistaffcosmo}) may well not be positive, and hence its
square root cannot be taken. A similar problem arose in the
original canonical FRW model and there it was interpreted as a
sign that the time choice $t:=\phi$ can only be defined in a
limited region of the reduced classical state space \cite{BI}.

We shall return to this issue later, but setting it aside for the
moment, our first step is to seek a `suitable' unitary
representation of the history group $C(T,
\mb{R}\times_{S}\mb{R}_+)$ with Lie algebra
(\ref{RWHACR1})--(\ref{RWHACR2}). By analogy with the discussion
in Section \ref{ACRrepn}, we see that the simplest representation
is when the Hilbert space $\mc{H}$ is the space of
square-integrable functions $\psi(t,r), (r>0)$ with measure $dt
dr/r$, and with operators acting as
            \beqa
&&(\hat r(s) \psi)(t,r):=\delta(s,t)r\psi(t,r)\label{eq:hatr(s)}\\
&& (\hat \p(s)\psi)(t,r):=-i\delta(s,t)r \frac{\partial \psi
(t,r)}{\partial r} \label{eq:hatp(s)}\eeqa The (better defined)
smeared version is \beqa
&&(\hat r(f) \psi)(t,r):=f(t)\,r\psi(t,r)\\
&& (\hat \p(g)\psi)(t,r):=-ig(t)\,r \frac{\partial \psi
(t,r)}{\partial r}. \eeqa

However, this basic representation on $\mc{H}$ is inadequate for
two reasons:
\begin{enumerate}
\item[(i)] It corresponds to a history that is only `active' at a
single time-point $t\in T$, whereas we want histories at any
finite set of time points $t_1,t_2,\ld,t_N$ for all $N=1,2, \ld$.
For each fixed $N$ we can do this by taking the (symmetrised)
$N$-fold tensor product of $\mc{H}$ with itself, and then we could
take the direct sum of these to give the exponential Hilbert space
$\mc H$. However, this turns out to be a very reducible
representation of the history algebra, whereas we want one that is
irreducible. Adding a cocycle contribution to the action on $\exp
\mc{H}$ is one possible way of achieving this.

\item[(ii)] If we compute the action of the operator product $\hat
r(s) \hat r(u)$ in $\mc H$ we find that
            \beq
(\hat r(s)\hat r(u) \psi)(t,r)=\delta (s,u)\delta(s,t)r^2
\psi(t,r) \eeq and a similar result for the product $\hat \p
(s)\hat \p(u)$. However, the history Hamiltonian in which we are
 interested is (\ref{Hhistaffcosmo}) i.e.
 \beq \label{Hhistaffcosmo1}
\hat H^2(t)=\frac{1}{12}\hat \P^2(t)+12K \hat R^4(t)-\hat
R^6(t)t^2m^2 \eeq and it is clear that we are going to have
problems while trying to define such singular products of
operators at the same time point.
\end{enumerate}

In the case of normal quantum field theory, we know that the
representation of the CCR has to be chosen precisely in order that
the Hamiltonian should exist, and we might conjecture that the
same is true here. In HPO  theory we have seen that the Lie
algebra (be it canonical or affine) arises on an $n$-fold tensor
product $\mc{V}_n=\mc{H}_n \ot \mc{H}_{n-1} \ot \cd \mc{H}_1$ of
$n$-copies of the Hilbert space $\mc H$ of the canonical (resp.\
affine) theory. For that reason, in the original papers on
continuous time, an exponential Hilbert space construction was
used as this can be interpreted as a continuous tensor product of
copies of the Hilbert space $L^2(\mb{R})$ where we have
$\mc{V}_{\mr{cts}}=\ot_{t}L_t^2 (\mb{R}):= \exp L^2 (\mb{R},dt)$
\cite{CJI3}.

It seems natural therefore to use an exponential Hilbert space in
continuous-time history theory. In particular, we wonder if we can
use the \emph{cocycle--exponential} Hilbert space construction in
order to find  a representation of the history affine algebra of
our quantum cosmology model in which (\ref{Hhistaffcosmo1}) exists
as a well-defined operator.

\subsection{Exponential group representations and cocycles}
\label{ssect:ExpGrpRpn}

For the present work we shall only need a few basic definitions of
cocycles and exponential group representations and these are given
briefly here.

First, let $\mc{H}$ be a complex Hilbert space. Then the
exponential Hilbert space $\exp\mc{H}$ is defined by $\exp\mc{H}:=
\mb{C} \op \mc{H} \op (\mc{H}\ot_{\mr{Symm}}\mc{H})\op \cd =
\mb{C}\bop_{n=1}^{\8}\mr{Symm}^{n}\mc{H}$, where $\mr{Symm}$
denotes symmetrised tensor product  and where $\mb{C}$ is the
complex numbers. For the vectors $\f\in \mc{H}$, there are defined
the `exponential vectors'
            \beq
\exp\f:= 1\op\f\op\frac{1}{\sqrt{2}}\f\ot\f\op\frac{1}{\sqrt{3}}
\f\ot\f\ot\f\op\cd=\bop_{n=0}^{\8}\frac{1}{\sqrt{n}}(\ot\f)^{n}
\eeq A fundamental property of this construction is that the set
of all finite linear combinations of exponential vectors is dense
in $\exp\mc{H}$: \emph{i.e.}, any vector in $\exp\mc{H}$ can be
written as the strong limit of a sequence of finite linear
combinations of exponential vectors. Another key result is that
the inner products of $\exp\mc{H}$ and $\mc{H}$ are related by
            \beq
\langle \exp\f, \exp\psi
\rangle_{\exp\mc{H}}=e^{\langle\f,\psi\rangle_{\mc{H}}}.
  \eeq

Now let $g\ra \hat A_g$ be a unitary representation of a
topological group $G$ on the Hilbert space $\mc{H}$. The
\emph{cochains} of different degrees are defined by
            \beqa
 C^{0}(G, \mc{H})&:=&\mc{H}\\
 C^{q}(G, \mc{H})&:=&\{ \mr{continuous \ functions \ from} \
G\times G\times \ld \times G \  \mr{to} \ \mc{H} \}  , \ q \geq 1
\eeqa where the direct product $G\times G\times \ld \times G$ is
taken $q$ times. Then the coboundary operator is defined on
$C^{0}(G, \mc{H})$, $C^{1}(G, \mc{H})$ and $C^{2}(G, \mc{H})$ as
            \beqa
v \in C^{0}(G,\mc{H}), \  (\d v)(g)\!\!\!\! &:=&\!\!\!\! \hat A_g
v-v
\label{cocycle1} \\
f \in C^{1}(G, \mc{H}), \  (\d f)(g_1,g_2)\!\!\!\!&:=&\!\!\!\!
f(g_1)+ \hat A_{g_1}
f(g_2)-f(g_1 g_2)\\
f \in C^2 (G, \mc{H}), \ (\d f)(g_1, g_2, g_3)\!\!\!\!
&:=&\!\!\!\! \hat A_{g_1}f(g_1, g_3)-f(g_1 g_2, g_3)+ f(g_1, g_2
g_3)-f(g_1, g_2). \ \ \ \eeqa Using these coboundary operators,
the spaces of $1$-coboundaries and $1$-cocycles, and
$2$-coboundaries and $2$-cocycles are defined respectively as:
            \beqa
&& B^{1}(G,\mc{H}):=
 \{ \b \ :  G\ra \mc{H} \mid
\b(g)=\hat A_{g}v-v \}\label{cocycle2}\\
&& Z^1(G, \mc{H}) :=
 \{ \b \ : G \ra \mc{H} \mid \b(g_1 g_2 )= \b (g_1)+\hat A_{g_1} \b(g_2) \}\\
 && B^2(G, \mc{H}):=
\{f \ :   G\times G \ \ra
 \mc{H} \mid f(g_1, g_2) = \b(g_1)+\hat A_{g_1}\b(g_2)-
\b(g_1 g_2) , \ \b \:  G \ra  \mc{H}\} \ \ \ \ \  \ \\
&& Z^{2}(G, \mc{H}) := \{f
\: G\times G\ra \mc{H}
 \mid f(g_1, g_2)+f(g_1g_2, g_3)=f(g_1,
g_2 g_3)+\hat A_{g_1}f(g_2, g_3) \} . \ \ \ \  \eeqa

One can easily prove that every coboundary is a cocycle. However,
it is not always the case that every cocycle is a coboundary. But
\emph{if} it is, then our calculations are much simplified. Since
that latter case leads to representations that agree with
Klauder's results \cite{Kla70a70b73a73b} and \cite{KlaRep} we
shall in the present article investigate this subcase only. Note
however, that we retain the option to follow Klauder's example and
choose coboundaries that are singular in some way, so that in
effect we do get a non-trivial cohomological situation. Also, as
we shall see, an addition of a coboundary to a representation has
a marked effect on the regularisation of certain operators that we
have to do to get a complete theory.

Let $G$ be a topological group with a unitary representation
$g\mapsto \hat A_g$ of $G$ on $\cal H$, and let $\b$ be a
$1$-cocycle. In addition let $\l:G\ra U(1)$ be a map such that
            \beq
\l(g_1 g_2)=\l(g_1) \l(g_2) e^{-i\Im \langle\b(g_1), \hat
A(g_1)\b(g_2)\rangle} \eeq for all $g_1,g_2\in G$. Then the
associated \emph{exponential representation} of $G$ on
$\exp\mc{H}$ is the family of operators $\hat U_g$, $g\in G$,
defined by
                \beq
\hat U_g \exp v= \l(g)e^{-\frac{1}{2}\| \b(g)\|^2 - \langle\hat
A(g)v,\b(g)\rangle^*} \exp (\hat A(g)v+ \b(g)).\eeq It can be
shown that the operators $\hat U_g$ are unitary and that $\hat
U_{g_1}\hat U_{g_2}=\hat U_{g_1 g_2}$. Thus we do indeed have a
unitary representation of $G$.

Now suppose that $\b=\d\f_0$ for $\f_0 \in \mc{H}$ \emph{i.e.},
$\b(g)=\hat A_g\f_0 - \f_0$ (as  follows from (\ref{cocycle1}) and
(\ref{cocycle2})). Then the real function $(g_1,g_2)\mapsto
 \Im \langle\b(g_1), \hat A(g_1)\b(g_2)\rangle \in C^2(G, \mb{R})$ is the
 coboundary of $g\ra \Im \langle\hat A(g)\f_0,\f_0\rangle \in C^1 (G,\mb{R})$,
 and we can choose $\l(g)=e^{-i \Im \langle\hat A(g)\f_0, \f_0\rangle}$
 for all $g\in G$.
  Then
        \beq
\hat U_g \exp v= e^{-i\Im \langle\hat
A(g)\f_0,\f_0\rangle}e^{-\frac{1}{2} \langle\hat A(g)\f_0-
 \f_0, \hat A(g)\f_0-\f_0\rangle - \langle\hat A(g)v,\hat A(g)\f_0-
 \f_0\rangle^*}
\eeq {\em i.e.},
        \beq \label{cocyc-cobound-action}
\hat U_g \exp v= e^{\langle\f_0,(\hat
A(g)-\mb{1})(\f_0+v)\rangle}\exp (\hat A(g)(v+\f_0)-
 \f_0). \eeq

Also, if we define
            \beq
E_{v}(g)=\frac{\langle\exp v, \hat U_g \exp v\rangle}{\langle\exp
v,\exp v\rangle} \eeq then we find that
            \beq \label{expectcocycl}
E_{v}(g)= e^{\langle(\phi_0+v),(\hat A(g)-1)(\f_0+v)\rangle}. \eeq

\ssect{Application to the FRW model}

It is clear that equations (\ref{cocyc-cobound-action}) and
(\ref{expectcocycl}) can be used in our FRW model form with $\cal
H$ chosen to be $L^2(T\times\mb{R},dt dr/r)$, and with the
operators defined in (\ref{eq:hatr(s)}) and (\ref{eq:hatp(s)}). In
particular, we define $\hat A(f):= e^{i\hat r[f]}$, and the
associated exponential representation as being  denoted $\hat
U_f:= e^{i\hat R[f]}$, where $f$ is a smearing function.

The simplest case for our representation would be to calculate
(\ref{expectcocycl}) for the case $v=0$. The exponential vector
$\exp 0$ is a cyclic vector and, as a consequence any vector in
the Hilbert space can be obtained by taking (the limit of) linear
combinations of the form $\hat U_g \exp 0$, where $g$ ranges over
the elements of the affine history group. In practice, a key role
is played by the generating functional
            \beq
E(g)=\langle\exp 0, \hat U_g \exp 0\rangle= e^{\langle\f_0,(\hat
A(g)-1)\f_0\rangle} \eeq which for the FRW model case becomes
            \beq \label{expectR}
E_{R}(f)=\langle\exp 0, e^{i \hat R[f]} \exp
0\rangle=e^{\langle\f_0, (e^{i\hat r[f]} -1)\f_0\rangle} \eeq
where $f$ is a test function. After computing the functional
derivative $\d/\d f(s)$ of (\ref{expectR}) and then taking the
limit $f\ra 0$ we find
\begin{eqnarray}
\langle\exp0,\hat{R}(s)\exp0\rangle&=&\langle\phi_0,\hat{r}(s)\phi_0\rangle\\
&=& \int_{\mb{R}_+}|\phi_0(s,r)|^2dr
\end{eqnarray}
where, in computing the second line, we have used
(\ref{eq:hatr(s)}).

We now take the second derivative of (\ref{expectR}) in the limit
$f\ra 0$ and take into account the fact that, on the Hilbert space
$\mc{H}$, we have
\begin{equation} \label{rphi0}
(\hat r(s)\phi_0)(r,t)=  \delta (s-t)r \phi_0(r,t).
\end{equation}
Moreover, it is easy to show that $\hat r(s) \hat r(t)= \hat r(t)
\hat r(s)$ and therefore that $\hat R(s) \hat R(t)=\hat R(t) \hat
R(s)$. As a result we find
\begin{eqnarray}
 & & \langle \exp0,\hat{R}(s)\hat{R}(t)\exp0  \rangle=
\langle \phi_0,\hat{r}(s)\hat{r} (t)\phi_0 \rangle +\langle
\phi_0,\hat{r}(s)\phi_0 \rangle \langle \phi_0,\hat{r}(t)\phi_0
\rangle
\label{expecRR} \\
&&=\d(s-t) \int_{\mb{R}_+} |\phi_0(s,r)|^2r\,dr + \int_{\mb{R}_+}
|\phi_0(s,r_1)|^2dr_1 \int_{\mb{R}_+}|\phi_0(t,r_2)|^2dr_2.
\end{eqnarray}
Similarly
 \beqa \label{expectP}
 E_{\P}(f):=\langle\exp 0, e^{i \hat
\P[f]} \exp 0\rangle=e^{\langle\f_0, (e^{i\hat \p[f]}
-1)\f_0\rangle} \eeqa and we find
\begin{eqnarray}
\langle\exp0,\hat{\P}(s)\exp0\rangle&=&\langle\phi_0,\hat{\p}(s)\phi_0\rangle\\
&=&-i\int_{\mb{R}_+}\phi_0(s,r)^*{\partial\phi_0(s,r)\over dr}dr
\end{eqnarray}
and
\begin{equation}
\langle\exp0,\hat{\P}(s)\hat{\P}(t)\exp0\rangle=
\langle\phi_0,\hat{\p}(s)\hat{\p}(t)\phi_0\rangle +
\langle\phi_0,\hat{\p}(s)\phi_0\rangle\langle\phi_0,\hat{\p}(t)\phi_0\rangle.
\label{expecPP}
\end{equation}

\sssect{The action on $\exp 0$} Moving beyond expectation values
we note that (\ref{cocyc-cobound-action}) with $v=0$ gives
            \beq
\hat U_g \exp0=e^{\langle\f_0, (\hat A_g-1)\f_0\rangle}\exp\{
(\hat A_g-1)\f_0\} \eeq which for the FRW model case becomes
            \beq \label{eexp0}
e^{i\hat R[f]}\exp0 = e^{\langle\f_0,(e^{i\hat
r[f]}-1)\f_0\rangle}\exp \{( e^{i \hat r[f]}-1)\f_0 \}. \eeq
 After we perform exponential expansion and  take the
  limit $\frac{\d}{\d f(s)} | _{f\ra 0} $ we find:
\begin{equation}
\hat{R}(s)\exp0 = \langle\phi_0,\hat r(s)\phi_0\rangle\exp0 + \hat
r(s)\phi_0
\end{equation}
where the `$+$' on the right hand side refers to the sum in the
exponential Hilbert space: {\em i.e.}, the direct sum of the
symmetrised tensor products.

 After taking the second derivative of (\ref{eexp0}), in the limit
$f\ra 0$ we find
\begin{eqnarray}
 \hat R(s) \hat R(t)\exp0 &=&
  \langle\phi_0,\hat r(s) \hat r(t)
\phi_0\rangle \exp 0 \nonumber +\langle\phi_0,\hat r(s)
\phi_0\rangle\langle\phi_0,\hat r(t)\phi_0\rangle\ \exp
0 \nonumber \\
&+& \langle\phi_0,\hat r(s)\phi_0\rangle\hat
r(t)\phi_0+\langle\phi_0,\hat r(t)\phi_0\rangle\hat r(s)\phi_0 +
\hat r(s)\hat
r(t)\phi_0\nonumber \\
&+& \frac{1}{\sqrt{2}} \{ \hat r(s)\phi_0\otimes \hat r(t) \phi_0
+ \hat r(t)\phi_0\otimes \hat r(s) \phi_0 \} \label{RRexp0} .
\end{eqnarray}
Similarly, we find
\begin{eqnarray}
 \hat \Pi(s) \hat \Pi(t) \exp0 &=& \langle\phi_0,\hat \p(s) \hat
\p(t) \phi_0\rangle \exp 0 \nonumber +\langle\phi_0,\hat \pi(s)
\phi_0\rangle\langle\phi_0,\hat \pi(t)\phi_0\rangle\ \exp 0
\nonumber \\
&+& \langle\phi_0,\hat \pi(s)\phi_0\rangle\hat
\pi(t)\phi_0+\langle\phi_0,\hat \pi(t)\phi_0\rangle\hat
\pi(s)\phi_0 + \hat \pi(s)\hat
\pi(t)\phi_0 \nonumber \\
&+& \frac{1}{\sqrt{2}} \{ \hat \pi(s)\phi_0\otimes \hat \pi(t)
\phi_0 + \hat \pi(t)\phi_0\otimes \hat \pi(s) \phi_0 \} .
\label{PPexp0}
\end{eqnarray}
where
\begin{equation} \label{pphi0}
(\hat \pi(s)\phi_0)(r,t)=-i\delta(s-t)r \frac{\partial
\phi_0}{\partial r}(r,t).
\end{equation}
 In a similar way we find
            \beqa
 \hat R(s)\hat \P(t)\exp 0 &=& \langle\f_0,\hat r(s)\hat
\p(t)\f_0\rangle \nonumber + \langle\f_0,\hat r(s)\f_0\rangle\hat
\p(t)\f_0 \nonumber \\ &+& \langle\f_0,\hat \p(t)\f_0\rangle\hat
r(s)\f_0+ \hat r(s)\hat \p(t)\f_0 \eeqa and similarly for $\hat
\P(\l) \hat R(s)$.

It is easy to check that the affine algebra of the operators $\hat
R(s)$ and $\hat \P(s)$ does indeed obey the classical Poisson
brackets, as it should; {\em i.e.},
\begin{eqnarray}
[\hat R(s), \hat \Pi(t)]\exp0&=&\hat R(s) \hat \Pi(t)\exp0-\hat
\Pi(t) \hat R(s)\exp0 \nonumber \\
&=& \langle\phi_0,[\hat r(s), \hat \pi(t)]\phi_0\rangle\exp0+
[\hat
r(s), \hat \pi(t)]\phi_0 \nonumber \\
&=& i \delta(s-t) \hat R(s) \exp0
\end{eqnarray}
and also by performing a third order expansion of the exponentials
it is easy to check that
\begin{equation}%\label{eq:RPP1}
[\hat R(s), \hat \Pi(t) \hat \Pi(m)]\exp 0 =i\delta(s-m) \hat
\Pi(t)\hat R(s)\exp 0 +i\delta(s-t) \hat R(s)\hat \Pi(m)\exp 0
\end{equation}
and that
\begin{equation}
[\hat \Pi(s), \hat R(t) \hat R(m)]\exp 0= - i\d(s-m)\hat R(t)\hat
R(m) \exp 0 - i\d(s-t)\hat R(t) \hat R(m) \exp 0
\end{equation} and that, indeed, all the classical Poisson
brackets between $\hat R(s)$ and $\hat \P(s)$ are respected.

From equations (\ref{expecPP}), (\ref{RRexp0}), (\ref{rphi0}),
(\ref{expecRR}), (\ref{pphi0}) and (\ref{PPexp0}) it is easy to
see that the quantities $\hat R(s)^2$ and $\hat \P(s)^2$ can
\emph{not} be evaluated within the present Hilbert space (and
therefore within the present exponential Hilbert space), just as
the square of the Dirac delta function can not be defined, or just
as the Dirac delta function with a zero argument does not have any
meaning (it diverges). This is true for any choice of the
coboundary, and therefore the need for some sort of regularisation
becomes apparent.

\sssect{The $\hat \t$ operator} \label{hatt}

A key ingredient of the proposed regularisation is the
introduction of the family of operators, $\hat\t(s)$, $s\in T$,
which are defined by
            \beq \label{taupsi}
(\hat \t (s)\psi)(t,r):=\d(s-t)\psi(t,r). \eeq In smeared form we
have \beq
    (\hat \t(f)\psi)(t,r):=f(t)\psi(t,r).\eeq

It is useful also to
 to introduce the operators $\hat r$ and
$\hat \p$ defined on $L^2(T\times \mb{R_+}; dtdr/r)$ as
            \beqa
&&(\hat r \psi)(t,r):=r \psi(t,r)\label{rpsi} \\
&&(\hat \p \psi)(t,r):=-ir\frac{\partial \psi(t,r)}{\partial r}
\label{pipsi} \eeqa which obey the ACR $[\hat r,\hat \p]=i \hat r
$. One can easily prove that, for all $s,s'\in T$,
            \beq \label{tautaupsi}
(\hat \t(s)\hat \t(s')\psi)(t,r)=\d(s-s')(\hat \t(s)\psi)(t,r),
\eeq
so that $\hat \t(s)\hat \t(s')=\d(s-s')\ \hat \t(s) $. We
also have the commutation relations
            \beqa
&& [\hat \t(s),\hat \t(s')]=0 \\
&& [\hat \t(s),\hat r]=0    \label{[t(s)r]}\\
&& [\hat \t(s), \hat \p]=0. \label{[t(s)p]}\eeqa The critical
point is that the new operator $\hat \t(s)$ is related to the
operators $\hat r(s)$ and $\hat \p(s)$ by
            \beqa
\hat r(s)&=&\hat \t(s) \hat r \label{rtr} \\
\hat \p(s)&=&\hat \t(s)\hat \p. \label{ptr} \eeqa
 This will be
used extensively in what follows.

\sssect{$\hat \t(\x)$ as a projection operator}

One can prove that the operator $\hat \t(s)$ is a type of
projection operator, as can intuitively be seen from
(\ref{tautaupsi}) or from the relation
        \beq \label{tautau}
\hat \t(s)\hat \t(s')=\d(s-s')\hat \t(s) . \eeq More precisely,
for $\hat \t(f)=\int \hat \t(s) f(s)ds$ and $f$ any test function,
we have, using equation (\ref{tautau}),
            \beq \label{tfh}
\hat \t(f) \hat \t(h) = \hat \t(fh) \eeq
which shows that
(\ref{tfh}) actually corresponds to a representation of the
\emph{ring} structure of the space of test functions
$C(T,\mb{R})$. Moreover, if $\x$ is any characteristic function of
a subset of $T$, then from (\ref{tfh}) we have
            \beq
\hat \t(\x)\hat \t(\x)=\hat \t(\x^2)=\hat \t(\x) \eeq which means
that $\hat \t(\x)$ is a genuine \emph{projection} operator.

It now makes sense to define the operator $\hat t$  by
        \beq
(\hat t \psi)(t,r):=t\psi(t,r), \eeq
which is self-adjoint on
$L^2(T\times \mb{R_+}; dtdr/r)$. Making use of the spectral
theorem we have that, for all $t$ in $T$ and all Borel functions
$f$,
    \beq
(f(\hat t) \psi)(t,r)=f(t)\psi(t,r)=(\hat \t(f)\psi)(t,r). \eeq In
other words,
        \beq
\hat \t(f)=f(\hat t) \eeq for all $t$ in $T$.

Note that it is clear from (\ref{tautau}) that the unsmeared
quantity $\hat\t(s)$ is not itself a projection operator since,
formally, we have
\beq\label{tau(s)tau(s)} \hat \t(s) \hat
\t(s)=\d(0)\hat \tau(s) \eeq
and, of course, the right hand side
of this expression is not well-defined because of the $\d(0)$
factor. However, motivated by ideas discussed at length by
Klauder, the form of (\ref{tau(s)tau(s)}) suggests that it might
be useful to define a \emph{regularised} form of the operator
product according to the definition\footnote{In order to keep the
dimensions matching up, we should really introduce a constant $d$
whose dimension is time, and then define
$\hat\t(t)\hat\t(t)|_{reg}:=d\hat\t(t)$ rather than
(\ref{tautaurreg}). However, in what follows we will ignore such
niceties. If necessary, the dimensions can always be corrected at
the end of the calculations.} \beq\label{tautaurreg} \hat
\t(t)\hat \t(t)|_{reg}:=\hat \t(t) \eeq for each $t$ in $T$. We
shall see the utility of this construction shortly.

\sssect{The operator $\hat p_t$}

The reducibility of the representation on $\cal H$ of the history
affine algebra can be understood in terms of (\ref{[t(s)r]}) and
(\ref{[t(s)p]}) which show that for all test functions $\hat f$, $
\hat \t(f)$ commutes with the operators $\hat r(s)$ and $\hat
\p(s)$.

To explore this situation further, let us assume for now on that
$T=\mb{R}$ for our history group $C(T,\mathbf{R}\times_s
\mathbf{R}_+)$ with Lie algebra (\ref{RWHACR1})--(\ref{RWHACR2}).
Then, one can define a conjugate operator $\hat p_{t}=-id/dt$
which satisfies
    \beq
[\hat t, \hat p_t]=i \eeq and which has the associated
one-parameter family of unitary operators
    \beq
\hat v(s):=e^{is\hat p_t} \eeq where $s\in \mb{R}$. We see at once
that, for all $s\in\mb{R}$,
    \beq
\hat v(s)\ \hat t \  \hat v(s)^{-1}=\hat t + s  \eeq and the
representation of the extended algebra on ${\cal H}\simeq
L^2(\mb{R}\times \mb{R_+}; dtdr/r)$
 is now \emph{irreducible}. The complete set of commutation
 relations that is associated with this irreducible representation is
            \beqa
&& [\hat r(t), \hat r(t')]=0\\
&& [\hat r(t), \hat \p(t')]=i\d(t-t')\hat r(t) \\
&& [\p (t), \p(t')]=0 \eeqa supplemented with
            \beqa
&& \hat v(s) \hat r(t) \hat v(s)^{-1} =\hat r(t+s) \\
&& \hat v(s) \hat \p(t) \hat v(s)^{-1}=\hat \p(t+s)  \eeqa for all
$s,t$ in $\mb{R}$.

\ssect{The regularisation procedure}
\sssect{The basic definition}

At the end of section \ref{hatt} we observed that the new operator
$\hat\t(s)$, $s\in\mb{R}$, is related to the operators $\hat r(s)$
and $\hat\p(s)$ by
     \beqa
\hat r(s)&=&\hat \t(s) \hat r \label{rtr2} \\
\hat \p(s)&=&\hat \t(s)\hat \p \label{ptr2}. \eeqa These
relations, together with $\hat\t(t)\hat \t(s)=\d(t-s)\hat\t(s)$,
give as a consequence that
    \beq
[\hat r(t), \hat\pi(s)]=[\hat r, \hat \p]\hat \t(t)\hat
\t(s)=i\hat r \d(t-s)\hat\t(t)=i\d(t-s)\hat r(t) \eeq so that, in
this particular representation, the `ultralocal' factor $\d(t-s)$
in the affine-history algebra comes from the product of $\hat
\t(t)$ with $\hat\t(s)$.

The definition (\ref{tautaurreg}) of the regularised product $\hat
\t(t)\hat \t(t)|_{reg}:=\hat \t(t)$ then  suggests that if $f$ is
any real-valued function of $\hat r$,  a regularised form of the
function $f(\hat r (t))$ of $\hat r(t)$ can be defined
by\footnote{If one wants to be very precise it is necessary to
take care of the dimensions with the add of the additional
dimensional constant $c$ mentioned earlier.}
    \beq
f(\hat r(t))_{reg}:=f(\hat r)\hat \t(t)\label{frreg} \eeq for each
$t\in \mb{R}$. Similarly,
    \beq \label{hpreg}
h(\hat\p(t))_{reg}:=h(\hat \p)\hat \t(t) \eeq for any function $h$
of $\hat p$. Here, $f(\hat r)$ and $h(\hat\p)$ are defined in the
usual way using the spectral theorem for self-adjoint operators.
As a result, these  regularised products have the property that
    \beq\label{fhreg}
[f(\hat r(t))_{reg},h(\hat\p(s))_{reg}]=[f(\hat r), h(\hat
\p)]\d(t-s)\hat \t(t).  \eeq

\sssect{The Hamiltonian for the case $m=0$ and $K=0$; action on
$\exp 0$} Let us start with the simplest case, which is when the
mass, $m$, and curvature, $K$, parameters both vanish. Then the
history Hamiltonian (squared) is formally
    \beq \label{H00}
\hat{H}^2 (t)=\frac{1}{12}\hat{\Pi}^2(t) \eeq and our task is to
give some mathematical meaning to this expression, and to its
integrals over regions of the time axis.

Note that, as we said earlier, a critical requirement for any
regularisation scheme is that the regularised form of the right
hand side of (\ref{H00}) must be a \emph{positive} self-adjoint
operator in order that the square root can be taken to give the
operator which we are actually seeking, namely $\hat H(t)$. Of
course, we might expect that the taking of this square root may
itself involve some sort of regularisation.

We will explore this matter for the exponential Hilbert space
quantisation discussed in Section \ref{ssect:ExpGrpRpn} and with a
coboundary $\phi_0$ that is a function of $t$ and $r$. Whether or
not the function $\phi_0$ is a proper element of
$L^2(\mb{R}\times{\bf R}_+,dtdr/r)$---and hence a genuine
coboundary---remains to be seen.

To see the type of regularisation scheme that is suggested, let us
begin with equation (\ref{PPexp0}) and note that, formally,
\begin{eqnarray}
\hat\Pi(t)^2\exp 0 &=&\langle\phi_0,\hat\p(t)^2\phi_0\rangle\exp 0
+\langle\phi_0,\hat\p(t)\phi_0\rangle^2\exp 0 \nonumber \\
&+&\langle\phi_0,\hat\p(t)\phi_0\rangle\hat\p(t)\phi_0+\hat\p(t)^2\phi_0
+{\sqrt{2}}\hat\p(t)\phi_0\otimes\hat\p(t)\phi_0
\end{eqnarray}
The only divergent terms here are
$\langle\phi_0,\hat\p(t)^2\phi_0\rangle$ and $\hat\p(t)^2\phi_0$,
and we can regularise these using the techniques introduced above
to give
\begin{equation}
\langle\phi_0,\hat\p(t)^2_{reg}\phi_0\rangle:=\langle\phi_0,
\hat\p^2\hat\t(t)\phi_0\rangle
\end{equation}
and
\begin{equation}\label{p(t)2refphi0}
\hat\p(t)^2_{reg}\phi_0=\hat\p^2\hat\t(t)\phi_0.
\end{equation}

This suggests the definition
\begin{eqnarray}
\hat\Pi(t)^2|_{reg}\exp 0
&:=&\langle\phi_0,\hat\p^2\hat\t(t)\phi_0\rangle\exp 0
+\langle\phi_0,\hat\p(t)\phi_0\rangle^2\exp 0 \nonumber \\
&+&\langle\phi_0,\hat\p(t)\phi_0\rangle\hat\p(t)\phi_0+\hat\p^2\hat\t(t)\phi_0
+{\sqrt{2}}\hat\p(t)\phi_0\otimes\hat\p(t)\phi_0 .
\end{eqnarray}
In particular, the expectation value of $\hat H(t)^2_{reg}$ in the
state $\exp 0$ is
 \beq\label{H200reg}
\langle\hat{H}^2(t)_{reg}\rangle =
\frac{1}{12}\left(\langle\phi_0,\hat{\p}^2\hat\t(t)\phi_0\rangle +
\langle\phi_0,\hat\p(t)\phi_0\rangle^2 \right). \eeq

Now, as remarked above, the positivity of (\ref{H200reg}) is a
necessary condition if $\langle\hat{H}^2(t)_{reg}\rangle$ is to be
the expectation value of a positive operator, and hence one whose
square root can be taken. However, we note that, since $\hat\p$ is
self-adjoint,
$\langle\phi_0,\hat{\p}^2\hat\t(t)\phi_0\rangle=\langle\hat\p\hat\phi_0,
\hat\p\hat\t(t)
\phi_0\rangle=\langle\hat\p\hat\phi_0,\hat\t(t)
\hat\p\phi_0\rangle$ where we have also used the fact that
$[\hat\p,\hat\t(t)]=0$. It follows that the right hand side of
(\ref{H200reg}) is non-negative provided that \beq
\langle\hat\p\hat\phi_0,\hat\t(t)\hat\p\phi_0\rangle\geq0. \eeq
However, we have that $(\hat\t(t)\phi_0)(s,r)=\d(t-s)\phi_0(s,r)$
and hence
\begin{equation}\label{<H00>reg}
\langle\hat\p\hat\phi_0,\hat\p\hat\t(t)\phi_0\rangle =\int_{{\bf
R}_+}(\hat\p\phi_0)^*(r,t) (\hat\p\phi_0)(r,t){dr\over r}
\end{equation}
and the right hand side is indeed greater or equal to zero since
$\hat\p$ is a self-adjoint operator on $L^2(\mb{R_+},dr/r)$. Thus
$\langle\hat{H}^2(t)_{reg}\rangle$ is greater than or equal to
zero for any choice of the coboundary $\phi_0$.

\sssect{The Hamiltonian for the case $m=0$ and $K=0$; action on
$\exp v$}

The results above involving $\exp 0$ are encouraging, but of
course the fact that the single matrix element $\langle\exp
0,\hat{H}^2(t)_{reg}\exp 0\rangle$ is non-negative is not
sufficient to guarantee that the same can be said for the operator
$\hat{H}^2(t)_{reg}$ as a whole. To do this, we need to study the
matrix elements $\langle\exp u,\hat{H}^2(t)_{reg}\exp v\rangle$
for arbitrary $u,v$ in $\cal H$; or, to be more precise, we need
to \emph{define} the regularised operator $\hat{H}^2(t)_{reg}$ by
giving regularised values for these matrix elements.

The calculations are more complicated than those just involving
$\exp 0$ but the basic results are easy to understand, and here we
will just give the main ones. One key result is that for any
$u,v\in\cal H$, we have
\begin{eqnarray}
\hat \Pi(s) \hat \Pi(t) \exp v &=& \langle(\phi_0+v),\hat \p(s)
\hat \p(t) (\phi_0+v)\rangle \exp v \nonumber \\
&+& \langle(\phi_0+v),\hat \pi(s)
(\phi_0+v)\rangle\langle(\phi_0+v),\hat \pi(t)(\phi_0+v)\rangle
\exp v
\nonumber \\
 &+& \langle(\phi_0+v),\hat \pi(s)(\phi_0+v)\rangle\hat
\pi(t)(\phi_0+v)\nonumber \\  &+& \langle(\phi_0+v),\hat
\pi(t)(\phi_0+v)\rangle\hat \pi(s)(\phi_0+v) + \hat \pi(s)\hat
\pi(t)(\phi_0+v) \nonumber \\
&+& \frac{1}{\sqrt{2}} \{ \hat \pi(s)(\phi_0+v)\otimes \hat \pi(t)
(\phi_0+v) + \hat \pi(t)(\phi_0+v)\otimes \hat \pi(s) (\phi_0+v)
\}  \ \ \ \ \ \  \label{PPexp01}
\end{eqnarray}
and also
\begin{eqnarray}\label{expuPPexpv}
{\langle\exp u,\hat\Pi(t)\hat\Pi(s)\exp v\rangle\over \langle\exp
u,\exp v\rangle}&=&
\langle(\phi_0 +u),\hat \p(t)\hat \p(s)(\phi_0+v)\rangle \nonumber\\
&+& \langle(\phi_0+u),\hat\p(t)(\phi_0+v)\rangle\langle(\phi_0+u),
\hat\p(s)(\phi_0+v)\rangle.
\end{eqnarray}
We see at once that the regularised matrix elements of
$\hat{H}^2(t)$ are (or, to be more precise, can consistently be
defined to be)
\begin{eqnarray}\label{expuH2regexpv}
{\langle\exp u,\hat{H}^2(t)_{reg}\exp v\rangle\over \langle\exp
u,\exp v\rangle}&=& {1\over 12}\left\{\langle(\phi_0 +u),\hat
\p^2\hat \t(t)(\phi_0+v)\rangle
\right.\nonumber\\
&+&\left.\strut\langle(\phi_0+u),\hat\p(t)(\phi_0+v)\rangle
\langle(\phi_0+u),\hat\p(t)(\phi_0+v)\rangle\right\}.
\end{eqnarray}
In particular, for any $u$ in ${\cal H}\simeq L^2(\mb{R}\times
\mb{R_+}; dtdr/r)$ we have
\begin{eqnarray}\label{expuH2regexpv2}
{\langle\exp u,\hat{H}^2(t)_{reg}\exp u\rangle\over \langle\exp
u,\exp u\rangle}&=& {1\over 12}\left\{\langle\hat\p(\phi_0
+u),\hat \t(t)\hat \p(\phi_0+u)\rangle
\strut\right.\nonumber\\
&+&\left.\strut\langle(\phi_0+u),\hat\p(t)(\phi_0+u)\rangle
\langle(\phi_0+u),\hat\p(t)(\phi_0+u)\rangle\right\} .
\end{eqnarray}
The analogue of the argument used above in the context of
(\ref{<H00>reg}) shows that the right hand side of
(\ref{expuH2regexpv}) is non-negative for any choice of coboundary
$\phi_0$. Thus, as desired, the regularised operator
$\hat{H}^2(t)_{reg}$ is non-negative. Clearly, the choice of
$\phi_0$ affects the `ground state' energy of the history system.

\sssect{The definition of $\hat H(t)$ for the case $m=0$ and
$K=0$}

We must consider how to construct a regularised operator $\hat
H_{reg}(t)$ from the given regularised operator
$\hat{H}^2(t)_{reg}$. As an example of the basic ideas, consider
on the starting space $\cal H$ the formal operator $\hat
h^2(t):={1\over 12}\hat\p(t)^2$ which is divergent, but whose
regularised form would be defined as \beq\label{h2(t)reg} \hat
h^2(t)_{reg}:={1\over 12}\hat\p^2\hat\t(t).\eeq

One cannot literally take the square root of the operator in
(\ref{h2(t)reg}) since there is no operator $\hat\n(t)$ such that
$\hat\n(t)^2=\hat\t(t)$. What is needed is to use the same
regularisation technique as before. In particular we note that
$\hat\p^2$ is a positive, self-adjoint operator on
$L^2(\mb{R}_+,dr/r)$ and hence it has a square root
$\sqrt{\hat\p^2}$. This suggests defining an operator \beq
    \hat h(t)_{reg}:={1\over \sqrt{12}}\sqrt{\hat\p^2}\,\hat\t(t)
\eeq on ${\cal H}=L^2(\mb{R}\times\mb{R}_+,dtdr/r)$. Clearly, we
have the relation \beq
    (\hat h(t)_{reg})^2_{reg}=\hat h^2(t)_{reg}.
\eeq

These ideas can be developed to construct an operator $\hat
H(t)_{reg}$ on $\exp\cal H$ starting with the operator $\hat
H(t)^2_{reg}$. The details of this will be given in a later paper
that will deal with the decoherence function in this history
quantum cosmology model.

 \sssect{The Hamiltonian for the case $m\neq0$ and
$K\neq0$; action on $\exp 0$}

For general values of the parameters $m$ and $K$, the calculations
are far more complicated.  The history Hamiltonian is formally
\beq \label{Hhistaffcosmo2} \hat H^2(t)=\frac{1}{12}\hat
\P^2(t)+12K \hat R^4(t)-\hat R^6(t)t^2m^2 \eeq and the obvious
problem here is that one sees immediately that this operator is
not necessarily positive for $m\neq 0$, and hence its square root
cannot be taken.

To investigate this matter let us start with the simplest
expectation value to study, which is
        \beq \label{expHmK}
\langle\exp 0,\hat{H}^2(t)\exp 0\rangle \ =
\frac{1}{12}\left(\langle \phi_0,\hat{\p}(t)^2\phi_0\rangle
+\langle \phi_0,\hat{\p}(t)\phi_0\rangle^2\right) + 12K A(t) -
m^2t^2 B(t) \eeq where $A(t):= \langle\exp0,\hat
R^4(t)\exp0\rangle$ and $B(t):=\langle\exp0,\hat
R^6(t)\exp0\rangle$. In computing these latter expressions we use
the basic formula (\ref{expectR}) which we functionally
differentiate the appropriate number of times. As a result we find
that the regularised expression of (\ref{expHmK}) has the form
        \beq \label{expHmKreg}
\langle\ \exp 0,\hat{H}^2(t)_{reg}\exp 0 \rangle =
\frac{1}{12}\left(\langle\ \phi_0,\hat \p^2  \hat \t(t)\phi_0
\rangle +\langle \ \phi_0, \hat{\p} \hat \t(t)\phi_0
\rangle^2\right) + 12K A(t)_{reg} -  m^2t^2 B(t)_{reg} \ \ \ \
\eeq where
        \beqa
A(t)_{reg}&=& \langle\f_0,\hat r^4 \hat \t(t)\f_0\rangle+
\langle\f_0,\hat r \hat \t(t)\f_0\rangle^4 \nonumber + 4
\langle\f_0,\hat r \hat \t(t)\f_0\rangle\langle\f_0,\hat r^3
\hat \t(t) \f_0\rangle \\
&+&    3 \langle\f_0, \hat r^2 \hat \t(t)\f_0\rangle^2 + 6
\langle\f_0, \hat r \hat \t(t) \f_0\rangle\langle\f_0, \hat r^2
\hat \t(t) \f_0\rangle\langle\f_0, \hat r \hat \t(t)\f_0\rangle
\eeqa and
        \beqa\label{B(t)reg}
B(t)_{reg}&=&  \langle\f_0,\hat r^6 \hat \t(t)\f_0\rangle+
\langle\f_0, \hat r \hat \t(t)\f_0\rangle^6 + 10\langle\f_0,\hat
r^3 \hat \t(t) \f_0\rangle^2 + 15\langle\f_0, \hat
r^2\hat \t(t)\f_0\rangle^3 \nonumber\\
&+&  6 \langle\f_0, \hat r \hat \t(t)\f_0\rangle\langle\f_0, \hat
r^5 \hat \t(t) \f_0\rangle+ 15\langle\f_0, \hat r^2 \hat
\t(t)\f_0\rangle\langle\f_0, \hat r^4 \hat \t(t) \f_0\rangle
\nonumber \\
 &+&  15\langle\f_0, \hat r^4 \hat \t(t)\f_0\rangle\langle\f_0, \hat
r \hat \t(t)\f_0\rangle^2\nonumber +15\langle\f_0,\hat r^2 \hat
\t(t) \f_0\rangle\langle\f_0, \hat r \hat \t(t) \f_0\rangle^4 \\
&+& 60\langle\f_0, \hat r^3 \hat \t(t) \f_0\rangle\langle\f_0,
\hat r^2 \hat \t(t)\f_0\rangle\langle\f_0, \hat r \hat
\t(t)\f_0\rangle \nonumber
\\ &+& 16\langle\f_0,\hat r \hat \t(t) \f_0\rangle^2\langle\f_0, \hat r^3
\hat \t(t) \f_0\rangle+ 49\langle\f_0,\hat r^2 \hat \t(t)
\f_0\rangle^2\langle\f_0, \hat r \hat \t(t) \f_0\rangle^2 .  \eeqa

 After
using the definition (\ref{taupsi}) we arrive at a squared
regularised Hamiltonian whose precise form depends strongly on the
coboundary $\f_0(r,t)$.  We note that for some choices of the
function $\f_0(r,t)$ the quantities $A(t)$ and $B(t)$ will diverge
even though the coboundary is a member of $L^2(\mb{R}\times
\mb{R_+}; dtdr/r)$. As an illustration of a family of cocycles
where this does not happen consider $\f_0(r,t):=C
r^{n_1}e^{-ar^{n_2}-ibr^{n_3}}$, where $\{n_1, n_2, n_3\}$ are
integers, $\{a,b\}\in \mathbf{R}_+$ and $C$ is a (real)
 constant. We confirmed (with the aid of Maple) that
for different choices of $a, b, n_1, n_2, n_3$ and $C$ all the
parts of the Hamiltonian have different values, which means that
the Hamiltonian is very `cocycle dependant'.

This is important for several reasons. One reason is that the
negative contribution to the squared Hamiltonian can have
different values. This is important since the physical Hamiltonian
is the square root of the right hand side of (\ref{expHmK}) which
should, therefore, be a positive operator; however, its positivity
depends on the value of the last negative term. One relevant
remark here is that, unlike normal ordering, the
$\t$-regularisation technique we have employed is such as to
maintain the heuristic positivity of operators. Thus, for example,
it is clear the the right hand side of (\ref{B(t)reg}) is positive
for any $t$ and all choices of $\phi_0$ for which the relevant
integrals over $r$ are finite.

If our conclusions about using the `internal' time $t=\phi$ were
similar in history theory to the analogous one that arises in the
normal canonical scheme, we would say that the formalism can only
be used in situations where $t$ is sufficiently small that the
matrix elements of $\hat H^2(t)_{reg}$ are positive. To discuss
this in the present situation let us use the following example.\\

\noindent{\textbf{Examples}}

We will take some specific examples for several functions of the
cocycle to see what the above expectation value of the squared
Hamiltonian become in practice; in particular we shall see how the
negative term of (\ref{expHmKreg}) appears and for which functions
$\f_0$ it can take a `small enough' value so as to guarantee the
positivity of the quantity (if any) whose square root we wish to
take.

First we integrate (\ref{expHmKreg}) over $t$ using
(\ref{taupsi}). This gives (for real-valued cocycles)
    \beqa \label{H2reg00int}
\langle\ \exp 0,\hat{H}^2(t)_{reg}\exp 0 \rangle \!\!\!&=&\!\!\! -
\frac{1}{12}\int_{0}^{\infty}\f_0(r,t)\Big[\frac{\partial}{\partial
r}\f_0(r,t)+r\frac{\partial^2}{\partial r^2}\f_0(r,t)\Big]dr \nonumber \\
\!\!\!&-&\!\!\!
\frac{1}{12}\left[\int_{0}^{\infty}\f_0(r,t)\frac{\partial}{\partial
r}\f_0(r,t)dr\right]^2+12KA(t)-m^2t^2B(t)\ \ \ \ \ \eeqa where
    \beqa
A(t)&=&\int_{0}^{\infty}\f_0(r,t)^2r^3dr+3\left[\,\int_{0}^{\infty}
\f_0(r,t)^2 r dr\right]^2+4 \int_{0}^{\infty}
\f_0(r,t)^2dr\int_{0}^{\infty}\f_0(r,t)^2 r^2 dr \nonumber \\ &+&
6 \left[\int_{0}^{\infty}\f_0(r,t)^2 dr\right]^2
\int_{0}^{\infty}\f_0(r,t)^2 r dr +
\left[\int_{0}^{\infty}\f_0(r,t)^2 dr\right]^4 \eeqa and there is
an analogous expression for $B(t)$.

As a demonstration of how the choice of a cocycle affects the
results, let us take the simple example where $\f_0(r,t):=C r
e^{-r^n}$, where $C$ is a (real)  constant, and $n$ is an
integer\footnote{We have made the simple choice
$\f_0(r,t)=Cre^{-r^n}$  as this lies in the domain of the
self-adjoint operator $\hat\p(f)$, and it makes the expectation
value $\langle\f_0, \hat \p(f) \f_0\rangle$ vanish, which
simplifies the calculations a little. Note that this particular
class of functions $\f_0$ is not square-integrable in $t$.
Nevertheless, it gives rise to meaningful expressions for the
regularised operator $\hat{H}^2(t)_{reg}$.}. In this case, with
the aid of Maple, we arrive at the conclusion that the range of
the possible values for $t$ increases with $n$ and with, for
example, $t^2<34.3/m^2$ (approximately) for $K=1$, $n=100$,
$C<10^{-2}$. We get physical results with all values $\{
-1,0,1\}$, although whether the results for $K=-1$ are physical
(i.e., give a range of $t$ values for which the left hand side of
(\ref{H2reg00int}) is positive) depends on the value of $n$,
unlike the ones for $K=1$ which are always acceptable.

What makes a big difference in the range of the $t$ values is the
value of the constant $a$ if we modify our example to one with
$\f_0(r,t):=C r e^{-ar^n}$. In this case big values of $a$ give a
much wider range for $t$: for example, for $a=10^6,\ n=1,\
K=-1,0,1,\ C<10^{-2}$ we find $t^2<5.3 \times 10^{32}/m^2$. It is
clear from these results that the range of $t$ over which the
expected value of the history FRW Hamiltonian makes sense is very
cocycle dependant.

 \sssect{The Hamiltonian for the case $m\neq0$ and
$K\neq0$; action on $\exp u$}

For the general case of $u\neq 0$ where $u$ is any function in
${\cal H}\simeq L^2(\mb{R}\times \mb{R_+}; dtdr/r)$ we can perform
calculations analogous to those in the previous section and we
arrive at the following expression:
    \beqa \label{uHregu}
{\langle\exp u,\hat{H}^2(t)_{reg}\exp u\rangle\over \langle\exp
u,\exp u\rangle}&=& \frac{1}{12}\left(\langle\ (\phi_0+u),\hat
\p^2 \hat \t(t)(\phi_0+u) \rangle +\langle \ (\phi_0+u), \hat{\p}
\hat \t(t)(\phi_0+u)
 \rangle^2\right)\nonumber \\
&+& 12K A_{u}(t)_{reg} - m^2t^2 B_{u}(t)_{reg} \eeqa
 where
 \beqa
A_{u}(t)_{reg}\!\!\!&=&\!\!\!  \langle(\f_0+u),\hat r^4 \hat
\t(t)(\f_0+u)\rangle+
\langle(\f_0+u), \hat r \hat \t(t)(\f_0+u)\rangle^4 \nonumber \\
\!\!\! &+& \!\!\! 4 \langle(\f_0+u),\hat r \hat \t(t)
(\f_0+u)\rangle\langle(\f_0+u),\hat r^3 \hat \t(t) (\f_0+u)\rangle
+ 3 \langle(\f_0+u), \hat r^2 \hat \t(t)(\f_0+u)\rangle^2
\nonumber\\
\!\!\! &+&\!\!\! 6 \langle(\f_0+u), \hat r \hat \t(t)
(\f_0+u)\rangle\langle(\f_0+u), \hat r^2 \hat \t(t)
(\f_0+u)\rangle\langle(\f_0+u), \hat r \hat \t(t) (\f_0+u)\rangle
\ \ \ \  \eeqa and
        \beqa
 B_{u}(t)_{reg}\!\!\! &=&\!\!\!  \langle(\f_0+u),\hat r^6 \hat
\t(t)(\f_0+u)\rangle+ \langle(\f_0+u), \hat r \hat
\t(t)(\f_0+u)\rangle^6 \nonumber \\ \!\!\! &+&\!\!\!
10\langle(\f_0+u),\hat r^3(t)(\f_0+u)\rangle^2 +
15\langle(\f_0+u), \hat
r^2 \hat \t(t)(\f_0+u)\rangle^3 \nonumber\\
\!\!\! &+& \!\!\! 6 \langle(\f_0+u), \hat r \hat
\t(t)(\f_0+u)\rangle\langle(\f_0+u), \hat r^5 \hat \t(t)
(\f_0+u)\rangle\nonumber \\\!\!\! &+&\!\!\!  15\langle(\f_0+u),
\hat r^2 \hat \t(t)(\f_0+u)\rangle\langle(\f_0+u), \hat r^4\hat
\t(t) (\f_0+u)\rangle
\nonumber \\
\!\!\!  &+& \!\!\!  15\langle(\f_0+u), \hat r^4 \hat
\t(t)(\f_0+u)\rangle\langle(\f_0+u),
 \hat r \hat \t(t)(\f_0+u)\rangle^2\nonumber \\ \!\!\!  &+& \!\!\!
 60\langle(\f_0+u),
\hat r^3 \hat \t(t) (\f_0+u)\rangle\langle(\f_0+u), \hat r^2 \hat
\t(t) (\f_0+u)\rangle\langle(\f_0+u), \hat r \hat \t(t)
(\f_0+u)\rangle \nonumber
\\ \!\!\!  &+& \!\!\!  16\langle(\f_0+u),\hat r \hat \t(t)
(\f_0+u)\rangle^2\langle(\f_0+u), \hat r^3 \hat
\t(t)(\f_0+u)\rangle\nonumber \\ \!\!\! &+& \!\!\!
49\langle(\f_0+u),\hat r^2 \hat \t(t)
(\f_0+u)\rangle^2\langle(\f_0+u), \hat r \hat \t(t)
 (\f_0+u)\rangle^2 \nonumber \\
\!\!\! &+& \!\!\!  15\langle(\f_0+u),\hat r^2 \hat \t(t)
(\f_0+u)\rangle\langle(\f_0+u), \hat r \hat \t(t)
(\f_0+u)\rangle^4 . \eeqa We see that our conclusions will be of
essentially the same qualitative type as those in the previous
section where $u=0$.

Our first general conclusion is that, when working with cocycles,
the regularisation technique we have employed means that the
choice of cocyle has a significant effect on the final results. In
general, if we construct two cocycle representations of the
history algebra with cocycles that differ  by a coboundary, then
the two representations are unitarily equivalent. This means that
if we take a representation whose cocycle \emph{is} a coboundary,
then this representation is unitarily equivalent to the
representation in which the cocycle is not present at all (i.e.,
it is equal to zero) and one might expect therefore that nothing
new can be obtained by using only a coboundary. However, as we
have seen,  the form of the regularised squared-Hamiltonian
depends significantly on the coboundary. Thus specifying the
coboundary is an important ingredient in determining the final,
regularised form of the history Hamiltonian.

%Our second remark regards the results we gain by using the example
%of the cocycles as $\f_0(r,t):=C r e^{-a_{u}r^n}$ and with
%$u(r,t)$ chosen to be $u(r,t):=C r e^{-b_{u}r^m}$ that we used for
%just $\f_0$ in the previous section. As one may expect from
%comparing (\ref{expHmKreg}) with (\ref{uHregu}), the results we
%found in this general case are in line and analogous with the ones
%in the previous $u(r,t)=0$ case.

\newpage

\sect{Conclusions}

We have made a preliminary study of the history approach to
canonical quantum gravity for the model of an FRW universe coupled
to a scalar field $\phi$. We have taken an approach in which we
choose an internal time before quantisation rather than the other
alternatives in which, for example, one looks at a history
analogue of the Wheeler-DeWitt equation \cite{HW}. The particular
choice we have made is $t=\phi$ and we are able to compare our
history approach with the much earlier canonical study of this
situation \cite{BI}.

An important part of our programme was to use a history analogue
of the affine commutation relations for the metric field rather
than the normal canonical quantities. This is important as it
helps to guarantee that the quantised metric is indeed a metric.
However, this introduces many novel mathematical problems as the
representations of the history analogue of the affine commutators
are not at all known, whereas for normal canonical fields the
history analogues look like conventional free bosonic fields, and
the standard Fock space can be employed.

We started with the very singular representation of the history
ACR on $L^2(\mb{R}\times\mb{R};dt dr/r)$ in which only a single
time point is probed. We have extended this to an arbitrary finite
number of time points by going to an exponential Hilbert space
with a coboundary. Then we have introduced a Klauder-type
regularisation method for the (squared) Hamiltonian and shown how
the final regularised energy-squared depends on the choice of the
coboundary. It turned out that when $m\neq 0$, the  problem with
positivity of this operator does not arise for us quite as it did
in the old canonical quantisation scheme. In particular, we found
that there exist (many) choices for the cocycle which can give a
positive value for the expected squared Hamiltonian for a wide
range of $t$ values; different cocycle-choices give very different
results for the expectation value of the Hamiltonian.

Many interesting tasks await future research. Firstly it would be
interesting to look at representations with cocycles that are not
coboundaries to see how this affects our final results,
particularly the positivity of the operators. This involves some
very complicated calculations which could be done, for example, by
using Maple. Then we have to show how the square roots can be
taken for the massless case, and for the appropriate ranges of $t$
for the massive case. Finally, in this line we should look for
some more representations of the history affine algebra and try to
find a complete theory. This will be important when we try to
extend the scheme to more complicated minisuperspace models, or to
the full quantum gravity theory.

Then, beyond all this, we must study the decoherence functional
and look for interesting consistent sets of histories.

\bigskip

\bigskip

\noindent{\large \textbf{Acknowledgements}}

\bigskip

I want to thank Chris Isham for many very lengthy and very useful
discussions. I also want to thank Charles Wang for introducing me
to Maple programming and Peter McClintock for his careful reading
through the paper.

I want to thank the future that let this work happen.

% s.B. and s.B.

\end{document}